\documentclass{iopjournal}
\usepackage{graphicx} 
\usepackage{amsmath}
\usepackage[capitalise]{cleveref}
\usepackage{accents}
\usepackage{xcolor}
\usepackage{url}
\usepackage{hyperref}
\usepackage{dsfont}
\usepackage{placeins}
\usepackage{amssymb}
\usepackage{physics}
\usepackage{dsfont}

\let\expval\relax
\usepackage{qcircuit}

\usepackage[
  backend=biber,
  style=numeric-comp,
  sorting=none,
  maxnames=8,
  doi=false,
  url=true
]{biblatex}

\newcommand{\vecbra}[1]{\langle\mspace{-3mu}\langle #1 |}
\newcommand{\vecket}[1]{| #1 \rangle\mspace{-3mu}\rangle}
\newcommand{\vecbraket}[2]{\langle\mspace{-3mu}\langle #1 | #2 \rangle\mspace{-3mu}\rangle}

\newcommand{\nodagger}[0]{{\vphantom{\dagger}}}

\usepackage{comment}

\usepackage[dvipsnames]{xcolor}
\newcommand{\updated}[1]{{\textcolor{black}{#1}}}

\graphicspath{{figures/}}

\begin{document}

\title{Accelerating qubit reset through the Mpemba effect}

\author{Théo Lejeune$^{1}$\orcid{0000-0000-0000-0000}, Miha Papič$^{2,3}$\orcid{0000-0001-8695-0872}, John Goold$^{4}$\orcid{0000-0001-6702-1736}, Felix C. Binder$^{4}$\orcid{0000-0003-4483-5643}, Fran\c{c}ois Damanet$^{1}$\orcid{0000-0003-1699-0195}, Mattia Moroder$^{4,*}$\orcid{0000-0002-1046-9991}}

\affil{$^1$Institut de Physique Nucléaire, Atomique et de Spectroscopie, CESAM, Universit\'e de Li\`ege, Li\`ege, 4000, Belgium}

\affil{$^2$IQM Quantum Computers, Georg-Brauchle-Ring 23-25, 80992 Munich, Germany}

\affil{$^3$Department of Physics and Arnold Sommerfeld Center for Theoretical Physics,
Ludwig-Maximilians-Universität München, Theresienstr. 37, 80333 Munich, Germany}

\affil{$^4$School of Physics, Trinity College Dublin, College Green, Dublin 2, D02K8N4, Ireland}

\affil{$^*$Author to whom any correspondence should be addressed.}

\email{theo.lejeune@uliege.be, miha.papic@meetiqm.com, gooldj@tcd.ie, felix.binder@tcd.ie, fdamanet@uliege.be,  \\ moroderm@tcd.ie}

%\keywords{sample term, sample term, sample term}

\begin{abstract}
Passive qubit reset is a key primitive for quantum information processing, whereby qubits are initialized by allowing them to relax to their ground state through natural dissipation, without the need for active control or feedback.
However, passive reset occurs on timescales that are much longer than those of gate operations and measurements, making it a significant bottleneck for algorithmic execution.
Here, we show that this limitation can be overcome by exploiting the Mpemba effect, originally indicating the faster cooling of hot systems compared to cooler ones.
Focusing on the regime where coherence times exceed energy relaxation times ($T_2 > T_1$), we propose a simple protocol based on a single entangling two-qubit gate that converts local single-qubit coherences into fast-decaying global two-qubit coherences. This removes their overlap with the slowest decaying Liouvillian mode and enables a substantially faster relaxation to the ground state.
For realistic parameters, we find that our protocol can reduce reset times by up to $50\%$ compared to standard passive reset.
We analyze the robustness of the protocol under non-Markovian noise, imperfect coherent control and finite temperature, finding that the accelerated reset persists across a broad range of realistic error sources.
Finally, we present an experimental implementation of our protocol on an IQM superconducting quantum processor.
Our results demonstrate how Mpemba-like accelerated relaxation can be harnessed as a practical tool for fast and accurate qubit initialization.
\end{abstract}

\section{Introduction}
The ability to efficiently reset a qubit register into a reference state is one of the key building blocks of any quantum computer \cite{Divincenzo2000, Nielsen&Chuang2000}. To this aim, a number of qubit reset protocols have been devised and demonstrated. Broadly speaking, all of these protocols can be divided into two categories, which we refer to as \emph{active} and \emph{passive} reset techniques. Passive techniques rely on the natural relaxation of the qubit to its ground state due to its coupling to the environment. By waiting for several multiples of the qubit's inherent relaxation time, the qubit can be initialized to its ground state with high probability. This method is straightforward and does not require any additional hardware or control pulses and is also robust to leakage errors, i.e. errors where the state of the qubit leaves the computational subspace \cite{Marques2023, Lacroix2025}. On the other hand, active reset techniques involve the use of control pulses or feedback mechanisms to bring the qubit to its ground state \cite{Riste2012,Johnson2012,Govia2015, Valenzuela2006, Grajcar2008, Geerlings2013, Jin2015, Tan2017}. 

% moving to superconducting [missing citations]
For superconducting qubits, one of the leading platforms for scalable quantum computing, both active and passive reset techniques have been successfully implemented \cite{Riste2012, Johnson2012, Govia2015, Valenzuela2006, Grajcar2008, Reed2010, Geerlings2013, Jin2015, Tan2017, Marques2023, Lacroix2025}. The latter is typically the preferred choice due to its simplicity and reliability in combination with the limited coherence time of these systems \cite{IBMQuantum2025}. On the other hand, common active techniques on this platform involve measurement-based feedback \cite{Riste2012,Johnson2012,Govia2015} or using engineered dissipation by coupling the qubit to auxiliary dissipative on-chip elements \cite{Valenzuela2006, Grajcar2008, Reed2010, Geerlings2013, Jin2015, Tan2017, Egger2018, Magnard2018, Zhou2021-ze, Han2023}. While significantly faster compared to passive reset, measurement-based feedback methods require low-latency electronics to implement fast-feedback loops and additionally rely on high-fidelity qubit readout, which can be challenging to achieve in practice. More specifically, current processors have demonstrated median readout fidelities of approximately 99.0\%, which is significantly lower compared to both single- and two-qubit gates \cite{Google2019, IBM2023, IQM2024, Google2025, Tan2025}. Engineered dissipation based methods, on the other hand, can be implemented without the need for measurement and can also be used in the presence of leakage errors, but require the careful calibration of additional operations used to transfer the excitations from the qubit to the dissipative element \cite{Valenzuela2006, Grajcar2008, Reed2010, Geerlings2013, Jin2015, Tan2017, Marques2023, Aamir2025, Lacroix2025}.

% why mpemba 
Passive reset therefore remains a simple and robust method; however, some algorithms do not require all qubits to be measured at the end of each execution, most notably when probing local properties in the simulation of a larger quantum system \cite{Georgescu2014, fedorov2021, Miessen2023}. In this case, the qubits that are not measured might retain some quantum coherence, i.e. off-diagonal elements in the density matrix. If the passive reset timescale is computed based on the energy relaxation time $T_1$ only, the residual coherence in the qubit state will therefore lead to unwanted correlations between subsequent algorithm executions if the decay time of these off-diagonal elements $T_2$ is significantly longer than $T_1$. The regime $T_2>T_1$, which we consider throughout this article, is gaining prominence in superconducting hardware, notably due to material improvements~\cite{Bland2025}, the use of flux-tunable transmon qubits~\cite{Yan2016}, and the development of active error suppression techniques, such as dynamical decoupling~\cite{papic2023, Cywinski2008}. Note that the duration of a passive reset is always significantly longer compared to the execution time of the algorithm itself and therefore significantly affects the shot execution rate~\cite{wack2021}.  This is especially relevant in today's noisy intermediate-scale quantum (NISQ) era, where error mitigation techniques are required to suppress the noise of current devices at the cost of an exponential overhead in the number of required shots~\cite{IBM2023, Cai2023, Google2025Echoes}.

\begin{figure}[h]
    \centering
    \includegraphics[width=0.7\linewidth]{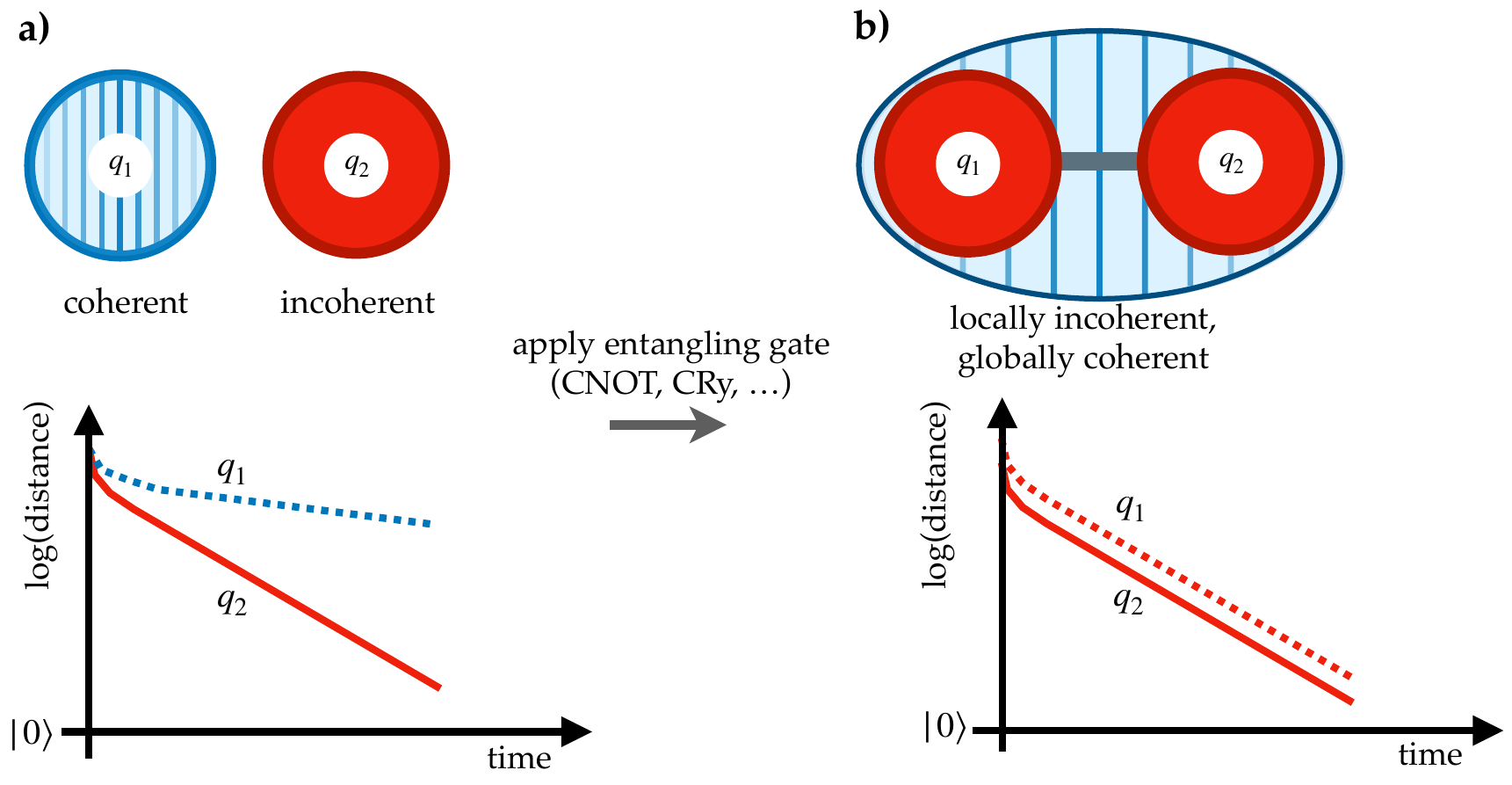}
    \caption{Enhancing qubit reset via coherence delocalization.
(a) In the regime $T_2 > T_1$, single-qubit coherences decay more slowly than populations, causing a coherent qubit $q_1$ (blue) to relax to the ground state more slowly than an incoherent qubit $q_2$ (red).
(b) Applying an entangling two-qubit gate (e.g., CNOT or CRy$(\pi)$, see~\cref{sec:cnot:protocol}) between $q_1$ and an incoherent ancilla converts local coherences of $q_1$ into global two-qubit coherences, which decay faster under local dissipation, thereby removing the slow relaxation bottleneck and accelerating the reset of $q_1$.}

    \label{fig:first:figure:cartoon}
\end{figure}

Hence, there is a need for the development of more advanced reset techniques to ensure proper qubit register initialization before each algorithm execution. One promising mechanism that can be exploited is the Mpemba effect, the counterintuitive phenomenon in which configurations initially further from equilibrium can relax faster~\cite{Lu2017, Klich2019, Kumar2020, Gal2020, Teza2023, Teza2025}.
Inspired by this effect, we propose a protocol to accelerate qubit reset in the more problematic regime when $T_2 > T_1$, which is illustrated in ~\cref{fig:first:figure:cartoon}.  
Our approach uses a single entangling gate between the target qubit and an incoherent ancilla that converts local single-qubit coherences into global two-qubit coherences, which decay much faster, thereby enabling a substantial speedup of passive reset without requiring any knowledge of the initial state.
Additionally, we show that our protocol is robust \updated{when the ancilla qubit is not fully incoherent} and even when taking into account a more realistic noise model, including non-Markovian (i.e., memory) effects for the qubit relaxation dynamics.

This article is organized as follows: 
First, in~\cref{sec:markovian:case}, we theoretically analyze the Mpemba-enhanced passive reset protocol, identifying conditions under which a speedup occurs.
We validate our protocol through numerical simulations on standard qubit models, showing its effectiveness on current quantum hardware.
Then, we assess its robustness by incorporating more realistic device error models, including non-Markovian noise in~\cref{sec:non-markovian:case} and imperfect control operations in~\cref{sec:imperfect:control}.
In~\cref{sec:experiment}, we present an experimental demonstration of the protocol on a superconducting quantum processor.
Finally, in~\cref{sec:conclusion:outlook}, we conclude by summarizing our main results and discussing their broader implications for quantum information processing.

\section{Accelerating qubit reset in the presence of Markovian dissipation}
\label{sec:markovian:case}
\subsection{The quantum Mpemba effect and the Davies map}
When a quantum system is weakly coupled to a Markovian (memoryless) bath, its reduced dynamics obey the Lindblad master equation~\cite{Lindblad1976},
\begin{equation}
    \frac{\mathrm{d} \hat{\rho}}{\mathrm{d}t} = \mathcal{L} [\hat{\rho}] = -\mathrm i [\hat{H}, \hat{\rho}] + \sum_l \underbrace{\left(\hat{L}^\nodagger_l \hat{\rho} \hat{L}^\dagger_l - \frac{1}{2} \{ \hat{L}^\dagger_l \hat{L}^\nodagger_l, \hat{\rho}\}\,\right)}_{\mathcal{D}_{\hat{L}_l}[\hat{\rho}]},
\label{eq:Lindblad}
\end{equation}
where $\mathcal{L}$ is the Lindbladian superoperator, $\hat{\rho}$ is the system density matrix, $\hat{H}$ its Hamiltonian, and the effect of the environment is encoded in the jump operators $\hat{L}_l$, which define the dissipator
$\mathcal{D}_{\hat{L}_l}$. 
Throughout this article, we will often employ vectorization \cite{AmShallem2015}, where density matrices are mapped to vectors (for instance via column stacking) $\hat{\rho}\to \vecket{\rho}$ and superoperators to operators $\mathcal{O} \to \hat{\mathcal{O}}$.
In this context, the Lindbladian takes the form
\begin{equation}
       \hat{\mathcal{L}} = -\mathrm i \hat{H} \otimes  \hat{\mathds{1}} +   \hat{\mathds{1}} \otimes \mathrm i\hat{H}^{\scriptscriptstyle T} + \sum_l \left(\hat{L}^{\nodagger}_l \otimes \left ( \hat{L}^{\dagger}_l \right ) ^{\scriptscriptstyle T} -\frac{1}{2} \hat{L}^{\dagger}_l \hat{L}^{\nodagger}_l \otimes  \hat{\mathds{1}}  - \frac{1}{2}  \hat{\mathds{1}} \otimes \left( \hat{L}^{\dagger}_l \hat{L}^{\nodagger}_l \right) ^{\scriptscriptstyle T}\right).
    \label{eq:Lindblad:vectorized}
\end{equation}
The spectral decomposition of the Lindbladian reads $\hat{\mathcal{L}} = \sum_k \lambda_k \vecket{r_k}\!\vecbra{l_k}$, where $\vecket{r_k}$ ($\vecket{l_k}$) denotes the right (left) eigenvector corresponding to the complex eigenvalue $\lambda_k$.
The eigenvalues have negative real parts, come in complex-conjugate pairs, and it is convenient to sort them as $\lambda_1 = 0 < \abs{\Re(\lambda_2)} \leq \abs{\Re(\lambda_3)} \leq \cdots$.
In the eigenbasis of $\hat{\mathcal{L}}$, the time evolution of the initial state $\vecket{\rho_i}$ can be written as~\cite{Carollo2021}
\begin{equation}
    \vecket{\rho(t)}=\mathrm e^{\hat{\mathcal L}t}\vecket{\rho_i}= \vecket{\rho_\mathrm{ss}}+\sum_{k=2}\mathrm e^{\lambda_k t}\vecbraket{l_k}{\rho_i}\vecket{r_k} \,,
    \label{eq:lindblad:decomposition}
\end{equation}
where the steady state $\vecket{\rho_\mathrm{ss}}$ is the right eigenvector $\vecket{r_1}$ corresponding to the eigenvalue $\lambda_1=0$.
\cref{eq:lindblad:decomposition} tells us that for an arbitrary initial state $\vecket{\rho_i}$, at late times the dynamics will be dominated by the slowest-decaying component, i.e., the equilibration speed will be proportional to $\exp(\abs{\Re(\lambda_2)}t)$. 
Instead, for a special initial state $\vecket{\rho_i'}$ having zero overlap with the slowest decaying mode $\vecket{l_2}$, the equilibration rate will be dictated by $\abs{\Re(\lambda_3)}$.
When such a fast-equilibrating state is initially further from the steady state with respect to some (pseudo-)distance function $D$, the curves defined by $D$ will cross in time, which is known as a strong Mpemba effect~\cite{Lu2017}.
\updated{In addition to its realization in Lindbladian dynamics~\cite{Nava2019, Carollo2021, Bao2022, Kochsiek2022, Ivander2023, Wang2024, Aharony2024, Moroder2024, , Xu2025, Longhi2025, Summer2025, Solanki2025, Xu2025, Beato2026}, a protocol has been developed for its unambiguous identification and classification~\cite{Nava2024}. Furthermore, the Mpemba effect has been investigated in non-Markovian open quantum systems~\cite{Strachan2024, Li2025, Zhang2025} and in the anomalous restoration of symmetries~\cite{Ares2023Nat, Ares2025, Rylands2024, Turkeshi2024, Liu2024,Joshi2024}. Another recent development concerns the discovery of the Pontus-Mpemba effect, which explicitly takes into account the time needed for preparing the system in the “further from equilibrium” initial state~\cite{Nava2025}.}
Moreover, its practical utility has recently been explored in the context of quantum state preparation~\cite{Westhoff2025} and the discharging of quantum batteries~\cite{Medina2024}.

An important type of Lindbladian is the Davies map~\cite{Davies1979}, which models the thermalization of a quantum system weakly coupled to a Markovian environment.
The dissipator is defined by jump operators corresponding to the transition elements of the system's Hamiltonian $\hat{H}$.
The associated prefactors satisfy the detailed balance condition, ensuring that the steady state is a thermal state.
Importantly, the vectorized generator $\hat{\mathcal{L}}$ of the Davies map can be brought into block-diagonal form, consisting of one smaller block that governs the evolution of the diagonal elements (populations) and a larger block that describes the dynamics of the off-diagonal elements (quantum coherences).
Furthermore, the real eigenvalues of $\hat{\mathcal{L}}$ correspond to populations, whereas complex eigenvalues are associated with coherences.
This separation between population and coherence dynamics provides a natural handle to selectively modify the relaxation behavior of a qubit by acting on its coherences alone~\cite{Moroder2024}.

\subsection{Turning local coherences into fast-decaying global coherences}
\label{sec:cnot:protocol}
\begin{figure}[!h]
    \centering
    % --- First row ---
    \begin{minipage}[b]{0.49\textwidth}
        \centering
        \includegraphics[width=\textwidth]{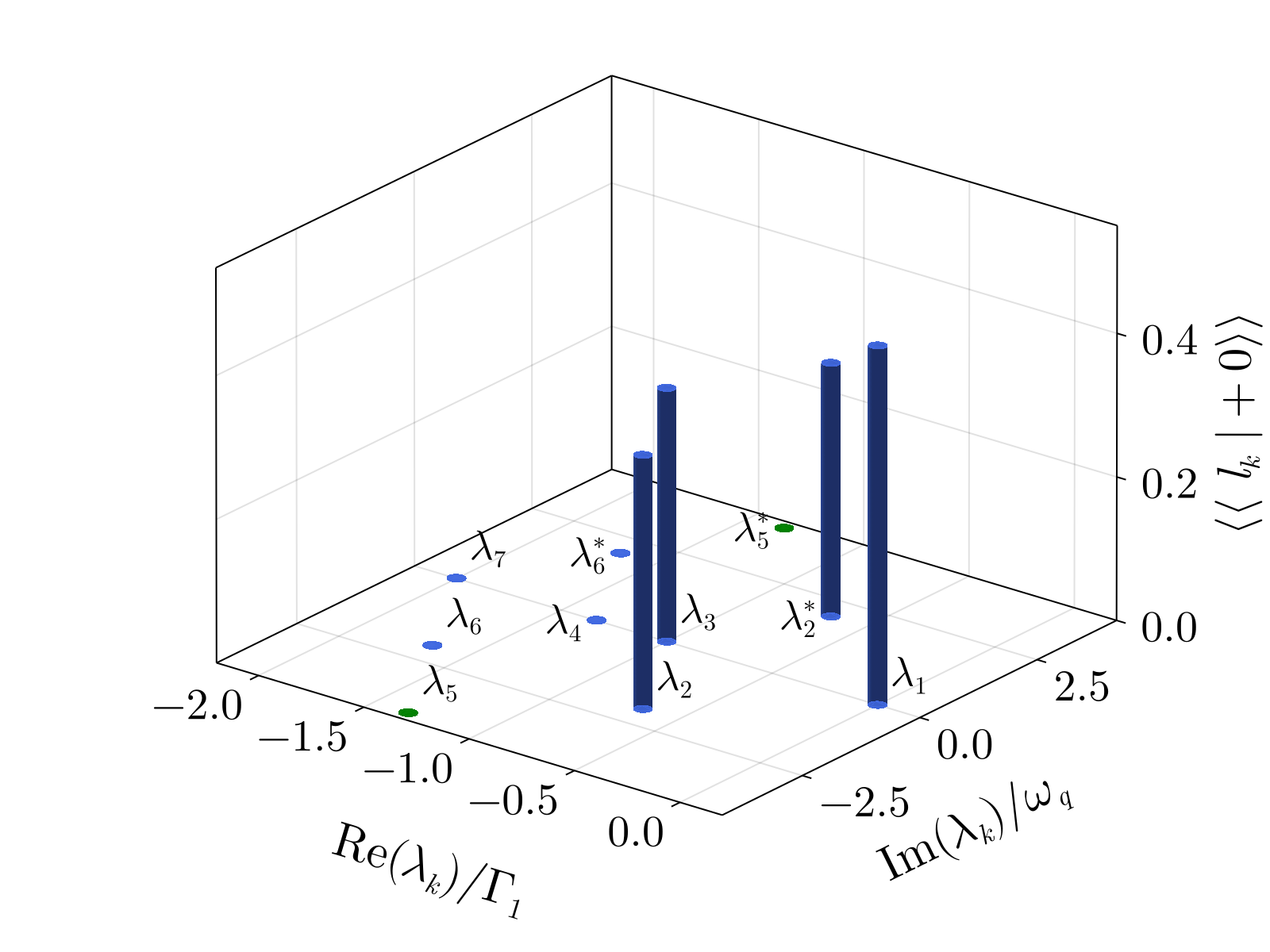}
        \textbf{(a)}
    \end{minipage}
    \hfill
    \begin{minipage}[b]{0.49\textwidth}
        \centering
        \includegraphics[width=\textwidth]{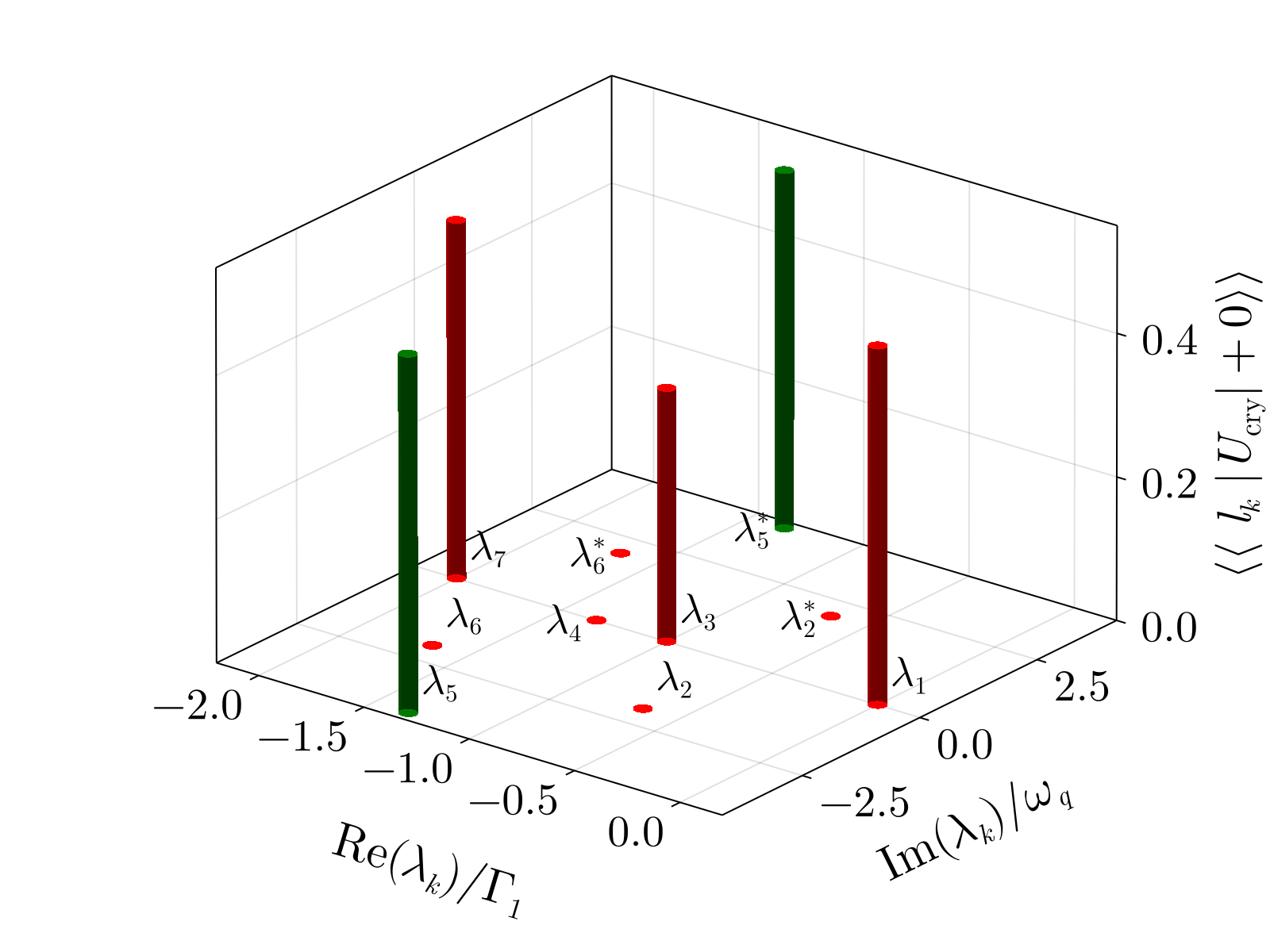}
        \textbf{(b)}
    \end{minipage}
    \vspace{0.2cm}
    % --- Second row ---
    \begin{minipage}[b]{0.49\textwidth}
        \centering
        \includegraphics[width=\textwidth]{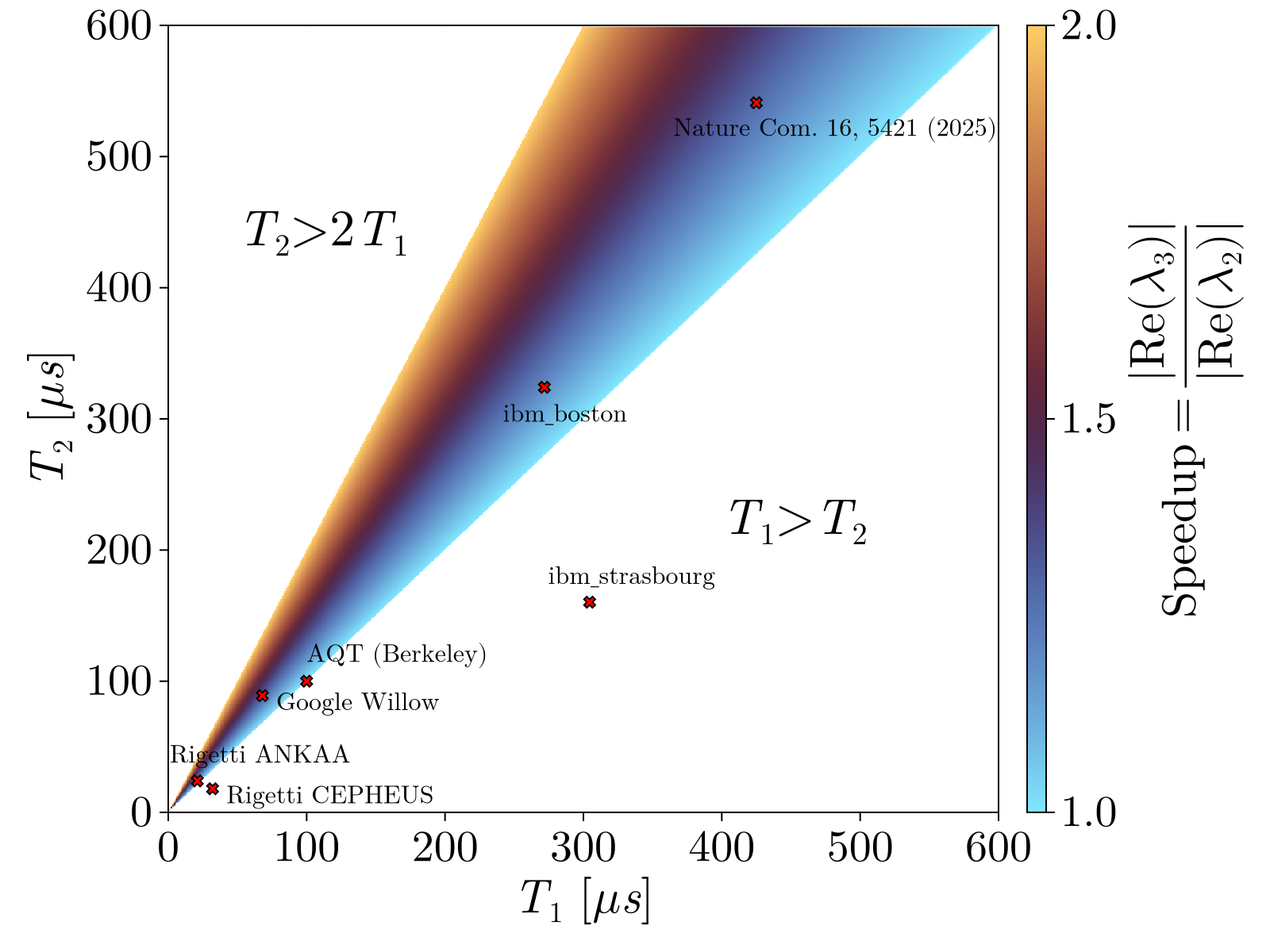}
        \textbf{(c)}
    \end{minipage}
    \hfill
    \begin{minipage}[b]{0.49\textwidth}
        \centering
        \includegraphics[width=\textwidth]{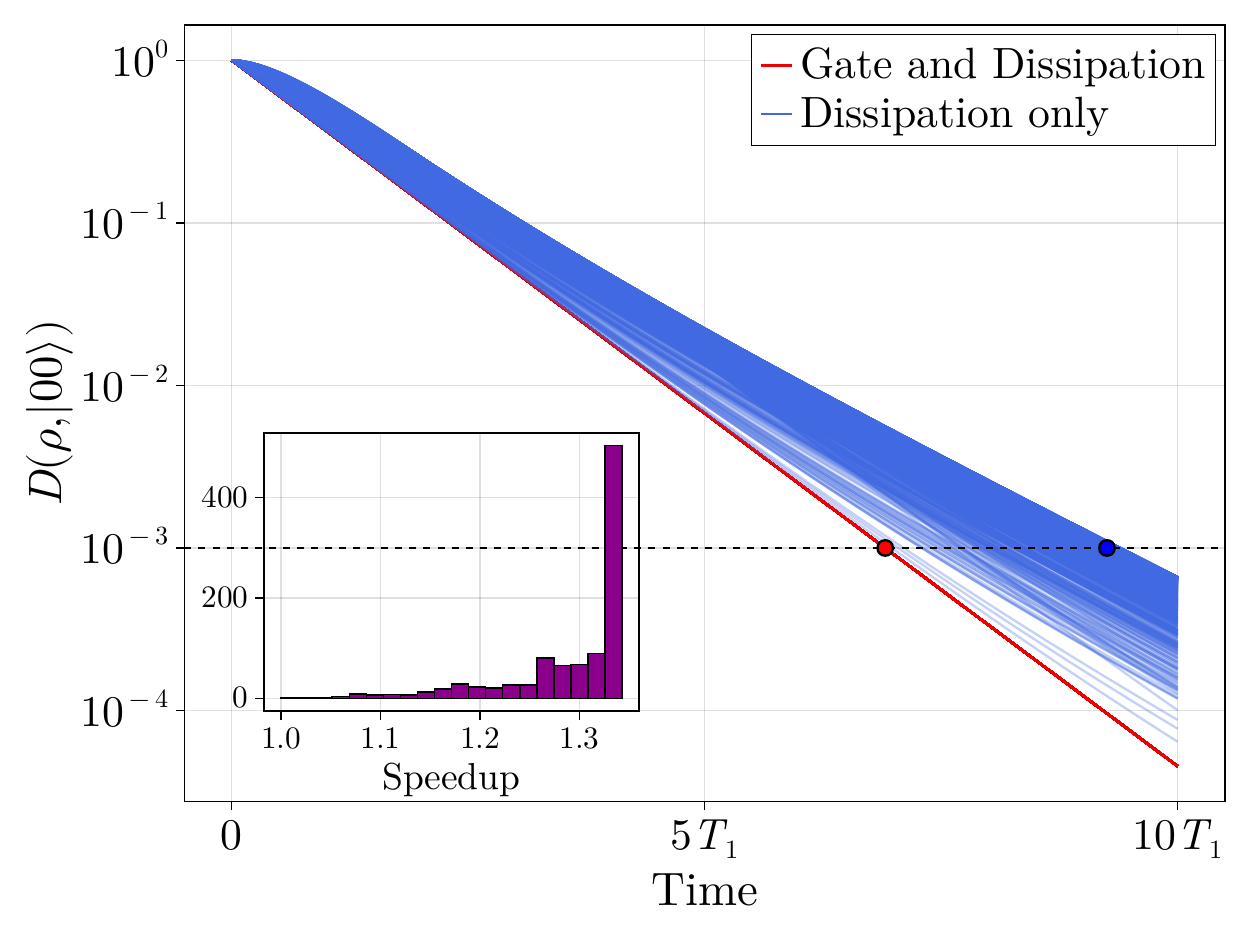}
        \textbf{(d)}
    \end{minipage}

    \caption{Speeding up qubit reset in the presence of Markovian noise. Panel (a) and (b) show the Liouvillian spectra from~\cref{eq:2qubit:davies:map:T0:with:dephasing}, along with the overlap $\vecbraket{l_k}{+0}$ between the initial state and the left eigenvectors before and after the application of the C-Ry gate, respectively. Panel (c) displays the asymptotic Mpemba speedup $|\mathrm{Re}(\lambda_3)|/|\mathrm{Re}(\lambda_2)|$ as a function of the $T_1$ and $T_2$ relaxation times. The red crosses represent the $T_1$ and $T_2$ values found in experimental setups. Finally, panel (d) shows the trace distance to the ground state as a function of time for $1000$ initial states, where the state of the system qubit $q_1$ is sampled from the Haar measure and the ancilla is initialized in the excited state ($p_2^1 = 1$). Results are shown with (red) and without (blue) application of the C-Ry gate.
    The inset shows the histogram of achieved speedups for $\epsilon=10^{-3}$ [see Eq.~(\ref{eq:speedup_gen})].}
    \label{fig:markovian}
\end{figure}

The spectral structure of the Davies map implies that, whenever
$\lambda_2$ is complex, the slowest relaxation mode is associated with the qubit coherences.
In this case, any operation that suppresses these coherences will accelerate relaxation toward the ground
state. Importantly, such suppression can be achieved \emph{without requiring any knowledge of the qubit’s
state}, by coupling the system qubit to an incoherent ancilla.
The central idea is to convert \emph{local} coherences of a qubit into \emph{global} coherences shared with an
ancilla qubit. While local coherences decay on a timescale set by the qubit dephasing rate, global two-qubit coherences decay faster under local dissipation, as they are affected by noise acting on either
qubit.

We consider a system qubit $q_1$ prepared in an arbitrary state
\begin{equation}
\hat{\rho}_1 =
\begin{pmatrix}
p_1^0 & C \\
C^*   & p_1^1
\label{eq:rho:1}
\end{pmatrix},
\qquad p_1^0 + p_1^1 = 1,
\end{equation}
\updated{where $C$ corresponds to \textit{local coherences} (i.e., only on qubit $1$),}
and an ancilla qubit $q_2$ prepared in an incoherent state
\begin{equation}
\hat{\rho}_2 =
p_2^0 \ket{0}\!\bra{0}
+
p_2^1 \ket{1}\!\bra{1},
\qquad p_2^0 + p_2^1 = 1.
\end{equation}
\updated{Interestingly, the following results extend to ancilla qubits with moderate residual coherences, as detailed in \cref{app:robustness:ancilla}, which implies the robustness of the protocol with respect to imperfect measurements.}
Then we consider a two-qubit unitary of controlled form
\begin{equation}
\hat U = \ket{0}\!\bra{0}\otimes \hat V_0 + \ket{1}\!\bra{1}\otimes \hat V_1 ,
\end{equation}
where $\hat V_0$ and $\hat V_1$ are single-qubit unitaries acting on $q_2$.
Applying $\hat U$ to the initial product state $\hat{\rho}_1\otimes\hat{\rho}_2$ yields a joint state
$\hat{\rho}'_{12}$ containing terms of the form
\begin{equation}
C\,\ket{0}\!\bra{1}\otimes \hat V_0 \hat{\rho}_2 \hat V_1^\dagger
\;+\;
C^*\,\ket{1}\!\bra{0}\otimes \hat V_1 \hat{\rho}_2 \hat V_0^\dagger ,
\end{equation}
which correspond, in the computational basis, to two-qubit coherences such as
$\ket{00}\!\bra{11}$. 
When tracing out the ancilla, the reduced state of
$q_1$ takes the form
\begin{equation}
\hat{\rho}_1' =
\begin{pmatrix}
p_1^0 & \kappa\, C \\
\kappa^*\, C^* & p_1^1
\end{pmatrix},
\qquad
\kappa = \mathrm{Tr}\!\left[\hat{\rho}_2\,\hat V_0^\dagger \hat V_1\right].
\end{equation}
Instead, when tracing out the system qubit, the reduced ancilla state becomes
\begin{equation}
\hat{\rho}_2'=\mathrm{Tr}_1[\hat{\rho}'_{12}]
= p_1^0\,\hat V_0\hat{\rho}_2\hat V_0^\dagger + p_1^1\,\hat V_1\hat{\rho}_2\hat V_1^\dagger .
\end{equation}
The terms proportional to the initial coherence $C$ reside in the off-diagonal blocks of $\hat{\rho}'_{12}$
(with respect to $q_1$) and therefore vanish under $\mathrm{Tr}_1[\cdot]$, so $C$ cannot generate local coherence on
$q_2$. Furthermore, if $\hat V_0,\hat V_1$ preserve diagonality, then $\hat{\rho}_2'$ remains incoherent. 
Thus, the action of the two-qubit gate on $q_1$ is equivalent to a pure dephasing channel, with local coherences proportional to $\kappa$. When $\kappa=0$, all local coherences of $q_1$ are removed after applying the gate.
The condition $\kappa=0$ depends only on the relative unitary
$\hat W = \hat V_0^\dagger \hat V_1$. Writing $\hat W$ as
\begin{equation}
\hat W = e^{i\phi}\!\left(
\cos(\theta/2)\,\hat{\mathds{1}} + i\sin(\theta/2)\,\hat n\cdot\vec{\hat\sigma}
\right),
\end{equation}
and expressing the incoherent ancilla as
$\hat{\rho}_2=\tfrac12(\hat{\mathds{1}} + r \hat Z)$ with $r=p_2^0-p_2^1$, one finds
\begin{equation}
\kappa = e^{i\phi}\big(\cos(\theta/2) + i\,r\,\sin(\theta/2)\,n_z\big).
\end{equation}
Requiring $\kappa=0$ for arbitrary incoherent ancilla populations implies $\cos(\theta/2)=0$ and $n_z=0$, so
that $\hat W$ is proportional (up to phase) to $\hat X$ or $\hat Y$. In this case, the gate suppresses local
coherences of $q_1$ independently of the state of the ancilla. If instead the ancilla is balanced
($p_2^0=p_2^1$), then any $\pi$ rotation, including $\hat Z$, suffices.

This condition is satisfied by a broad class of experimentally relevant two-qubit gates. In particular,
CNOT and controlled-$Y$ rotations naturally realize $\hat W\propto\hat X$ or $\hat Y$ and therefore suppress
local coherences for any incoherent ancilla. Native controlled-phase (CZ) gates, common in superconducting
and neutral-atom platforms, also realize the mechanism when the ancilla populations are balanced, or after
conversion to a CNOT via local single-qubit rotations. More generally, essentially all major quantum
computing platforms provide native entangling gates that are either directly of controlled form or can be
efficiently compiled into such a form.
In the remainder of this work, we focus mainly on the realization of
this mechanism based on a controlled $R_y(\pi)$ gate acting on an incoherent ancilla
\[
\Qcircuit @C=2.0em @R=2.0em {
    \lstick{\hat \rho_1}  & \ctrl{1} & \qw \\
    \lstick{\hat \rho_2} & \gate{R_y(\pi)} & \qw 
    } 
\]
For this choice, $\hat W\propto\hat Y$, and the condition $\kappa=0$ is satisfied for any incoherent ancilla
state. The gate therefore acts as a perfect dephasing channel on $q_1$, converting its local coherences into
purely global two-qubit coherences. For instance, considering $\hat{\rho}_\mathrm{tot} = \hat\rho_1 \otimes \ket{0}\!\bra{0}$ (with $\hat{\rho}_1$ defined in~\cref{eq:rho:1}) we obtain 
\begin{equation}
\hat{U}_\mathrm{cry}\hat\rho_\mathrm{tot}\hat{U}^\dagger_\mathrm{cry} = \begin{pmatrix}
        p_0 & 0 & 0 & C \\
        0 & 0 & 0 & 0 \\
        0 & 0 & 0 & 0 \\
        C^* & 0 & 0 & p_1 
    \end{pmatrix}.
    \label{rho_tot_CRY}
\end{equation}
\updated{The coherences $C$ of such state are referred to as \textit{global coherences}, since they correspond to the basis states $\ket{00}\!\bra{11}$ and $\ket{11}\!\bra{00}$ and are thus coherences of an entangled two-qubit state.}
In \cref{sec:qubit:dynamics} we compute the fast decay rate of these global coherences under local dissipation.
Finally, we note that the requirement of an incoherent ancilla is naturally satisfied in many practical
settings. In quantum algorithms where only part of the register is measured
\cite{IBM2023, Google2025Echoes}, the measurement process projects those qubits onto incoherent states.
These post-measurement qubits can therefore be reused as ancillas to suppress coherences in the remaining
qubits, enabling accelerated passive reset without additional control, feedback, or hardware overhead.

\subsection{Dissipative qubit dynamics} \label{sec:qubit:dynamics}

For idling superconducting qubits, the interaction with the environment is typically described by a zero-temperature Davies map with additional dephasing \cite{Burnett2019, papic2023,Cho_2023}
\begin{equation}
    \mathcal{L} [\hat{\rho}] = -\mathrm i \omega_q[\hat{\sigma}^{1}_z + \hat{\sigma}^{2}_z, \hat{\rho}] + \Gamma_1 \left(\mathcal{D}_{\hat{\sigma}^{1}_-}[\hat{\rho}] + \mathcal{D}_{\hat{\sigma}^{2}_-}[\hat{\rho}] \right) + \frac{\Gamma_\phi}{2}\left(\mathcal{D}_{\hat{\sigma}^{1}_z}[\hat{\rho}] + \mathcal{D}_{\hat{\sigma}^{2}_z}[\hat{\rho}]\right)\, .
\label{eq:2qubit:davies:map:T0:with:dephasing}
\end{equation}
Here, $\hat{O}^{1}=\hat{O} \otimes \hat{\mathds{1}}$ and $\hat{O}^{2} = \hat{\mathds{1}} \otimes \hat{O}$ are operators acting on $q_1$ and $q_2$ respectively. 
The first term of~\cref{eq:2qubit:davies:map:T0:with:dephasing} represents the unitary evolution of each qubit, at frequency $\omega_q$. 
The frequency detuning between idling qubits (which is present in real quantum hardware) can be safely neglected here, as the qubits evolve independently apart from the entangling gate.
The second term represents amplitude damping on both qubits with rate $\Gamma_1=1/T_1$ while the third describes pure dephasing with rate $\Gamma_\phi/2$. 
Note that both the second and the third term contribute to Markovian dephasing, characterized by the $T_2$ relaxation time as
\begin{equation}
    \Gamma_2 = \frac{1}{T_2} = \frac{\Gamma_1}{2} + \Gamma_\phi \, ,
    \label{eq:Markov_dephasing_rate}
\end{equation}
implying $T_2\leq2T_1$ for positive pure dephasing rates $\Gamma_\phi >0$.
In real superconducting hardware, temperature is small but finite, typically leading to a residual excited state population of order $1\%$.
We study the impact of a finite-temperature environment in \cref{app:finite:temperature}, showing that it does not negatively affect the performance of our protocol.

Before analyzing the time-dependent dynamics of the qubits described by \cref{eq:2qubit:davies:map:T0:with:dephasing}, it is insightful to see how the state of the system decomposes onto the eigenbasis of the Liouvillian before and after the application of the C-Ry gate.
The Liouvillian eigenvalues obtained from \cref{eq:2qubit:davies:map:T0:with:dephasing} are displayed in~\cref{fig:markovian}, along with the overlaps of the corresponding left eigenvectors $\vecbra{l_k}$ with an initial separable vectorized two-qubit state of the form $\vecket{\rho_i}
=
\bigl(\ket{+}\ket{+}\bigr)
\otimes
\bigl(\ket{0}\ket{0}\bigr)$, with $|+\rangle =(|0\rangle +|1\rangle)/\sqrt{2}$, before [panel (a)] and after [(panel (b)] applying the C-Ry gate (that is, using $\vecket{\rho_i}$ and $\vecket{\rho_i'}$, respectively). 
Note that here $\vecket{\rho_i}$ is represented using the Choi-Jamiołkowski isomorphism,
$\ket{\psi}\!\bra{\psi} \mapsto \ket{\psi}\ket{\psi}$.
It can be seen that the gate application suppresses the overlap of the state with the slowest decaying mode $\vecket{l_2}$, while increasing those with $\vecket{l_5}$ and $\vecket{l_7}$. Furthermore, the eigenvalues corresponding to the single-qubit coherences, $\ket{0}\!\bra{1}$ and $\ket{1}\!\bra{0}$, are $\lambda_2$ and $\lambda^*_2$ with real parts associated to a decay rate $(\Gamma_1/2)+\Gamma_\phi$. The two-qubit coherences $\ket{00}\!\bra{11}$ and $\ket{11}\!\bra{00}$ on the other hand, correspond to $\lambda_5$ and $\lambda^*_5$, represented by the green overlaps in \cref{fig:markovian}, and are associated to a decay rate $\Gamma_1+2\Gamma_\phi$, i.e. twice as large as the single-qubit ones.
Also, since $\lambda_3$ is real, the corresponding overlap is left untouched, so that $\vecket{l_3}$ becomes the slowest decaying mode of the modified initial state.

To quantify the relative reduction of relaxation time towards the steady state, we define the \textit{speedup} as 
\begin{equation}
    S(\epsilon) = \frac{t(\epsilon,\hat\rho_i)}{t(\epsilon,\hat\rho_i')},
    \label{eq:speedup_gen}
\end{equation}
where $t(\epsilon,\hat\rho) = \min\left\{t \big| D\left(e^{\mathcal{L} t}\hat\rho,\hat\rho_{\mathrm{ss}}\right) < \epsilon \right\}$ is the time needed, starting with an initial state $\hat\rho$, to reach the steady state $\hat\rho_{\mathrm{ss}}$ up to some distance $\epsilon$, where $D(\hat{A},\hat{B}) = \mathrm{Tr}[(\hat{A}-\hat{B})^\dagger(\hat{A}-\hat{B})]^{1/2}/2$ is the usual trace distance between any operators $\hat{A}$ and $\hat{B}$.
Since at long times we have that $D\left(e^{\mathcal{L} t}\hat\rho,\hat\rho_{ss}\right) \approx \vecbraket{l_j}{\rho_i} e^{\text{Re}(\lambda_j)t}$ with $j$ indexing the slowest decaying mode, \cref{eq:speedup_gen} yields~\cite{Westhoff_master_thesis}
\begin{equation}
    S(\epsilon) \approx \frac{\text{Re}(\lambda_3)}{\text{Re}(\lambda_2)}\frac{\ln(\epsilon) - \ln(\vecbraket{l_2}{\rho_i})}{\ln(\epsilon) - \ln(\vecbraket{l_3}{\rho_{i}'})} \, .
    \label{eq:speedup_full}
\end{equation}
The latter tends, for $\epsilon \ll 1$, to the asymptotic speedup
\begin{equation}
    S(\epsilon \to 0) = \frac{\mathrm{Re}(\lambda_3)}{\mathrm{Re}(\lambda_2)},
    \label{eq:asymptotic_speedup}
\end{equation}
which we represent in~\cref{fig:markovian} (c).
\updated{Note that~\cref{eq:speedup_full} holds with equality at late times, when the dynamics of $\vecket{\rho_i}$ and $\vecket{\rho_{i'}}$ are dominated by the slowest and second-slowest Lindblad modes, $\vecket{l_2}$ and $\vecket{l_3}$, respectively. Consequently, in the joint limit $t \to \infty$ and $\epsilon \to 0$, the asymptotic expression in~\cref{eq:asymptotic_speedup} becomes exact.}

Furthermore, we have
\begin{equation}
    \text{Re}(\lambda_2) = -\frac{\Gamma_1}{2} -  \Gamma_\phi \quad \text{and} \quad \text{Re}(\lambda_3) = -\Gamma_1\,.
    \label{eq:l2_l3}
\end{equation}
\updated{Then,~\cref{eq:Markov_dephasing_rate} and $\Gamma_1 = 1/T_1 $ together with \cref{eq:l2_l3} imply that $S(\epsilon \to 0) = T_2/T_1$.}
Also, this relation holds as long as $T_2\geq T_1$: once populations decay slower than coherences, applying the controlled Y rotation will not speed up the thermal relaxation, as shown in panel (c) of~\cref{fig:markovian}. \updated{Indeed, in such case the spectrum of the Liouvillian becomes such that the real parts of $\lambda_2$ and $\lambda_3$ of \cref{eq:l2_l3} are interchanged.}

The approach presented here applies to arbitrary coherent states, with the
achieved speedup depending on the initial state of $q_1$ and on the ratio $T_2/T_1$. To demonstrate this, we apply it to random pure states, i.e., randomly sampled on the surface of the Bloch sphere using the uniform Haar measure~\cite{Singh_2016}. 
The trace distances to the ground state of such states with and without application of a C-Ry gate, as a function of time, are displayed in panel (d) of ~\cref{fig:markovian} in red and blue, respectively. The panel's inset shows a histogram of the achieved speedups distribution, with a median speedup around $1.4$.
This follows from~\cref{eq:speedup_full}, which yields a speedup of $T_2/T_1$ (here set to $1.5$) multiplied by a finite $\epsilon$ correction at short times. 

We stress that while the analysis above was carried out with $\hat{\rho}_2=\ket{1}\!\bra{1}$, the effectiveness of the protocol does not rely on the specific choice of such a rank-deficient state.
As discussed in~\cref{sec:cnot:protocol}, the only requirement is that $q_2$ is prepared in an arbitrary incoherent state, such that the application of an entangling gate removes the local coherences of $q_1$ (and hence its overlap with the slowest-decaying Liouvillian modes).
Importantly, the rank of the initial state and its overlap with Liouvillian eigenmodes are independent notions, and generic full-rank states may still have vanishing overlap with specific decay modes.
We address this issue explicitly in~\cref{app:robustness:ancilla}, where we show that the protocol remains effective for arbitrary incoherent ancilla states.

\section{The effects of non-Markovianity}
\label{sec:non-markovian:case}

\begin{figure}[h]
    \centering
    \includegraphics[width=\textwidth]{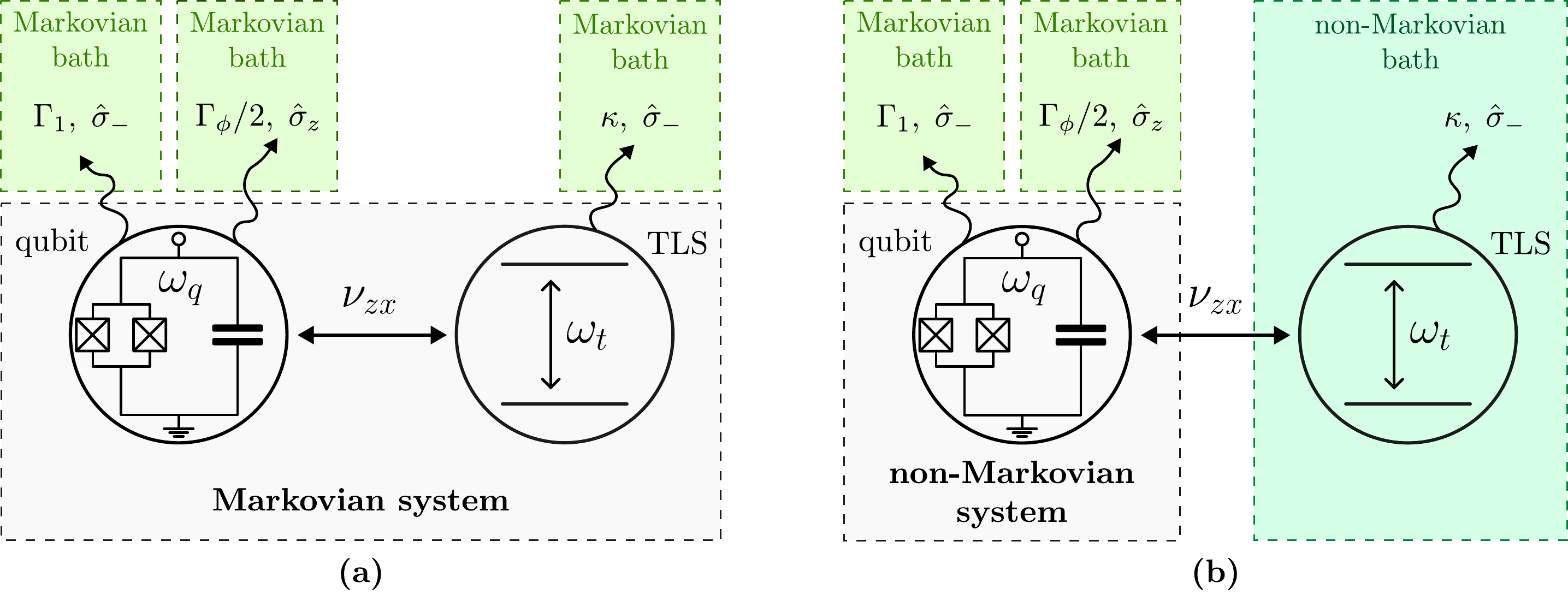}
    \caption{Markovian (a) and non-Markovian (b) descriptions of the system considered in \cref{sec:non-markovian:case}, where a flux-tunable transmon qubit of frequency $\omega_q$ is coupled, with strength $\nu_{zx}$, to a single TLS of frequency $\omega_t$. Both are subject to amplitude damping with rates $\Gamma_1$ and $\kappa$, respectively. Additionally, the qubit is subject to dephasing with a rate $\Gamma_\phi$/2. (a) Markovian description of the combined qubit-TLS system described by the Liouvillian $\mathcal{L}_{\mathrm{emb}}$ [Eq.~(\ref{eq:non-Markovian:model})]. (b) Non-Markovian description of the qubit, where the damped TLS acts as a non-Markovian bath with finite memory time $1/\kappa$. The reduced dynamics of the qubit obtained after tracing out the TLS is well described by the Redfield Liouvillian~$\mathcal{L}_{\mathrm{red}}$ [Eq.~(\ref{eq:reduced_liouv})].}
    \label{fig:NM_sketch}
\end{figure}

\begin{figure}[htbp]
    \centering
    % --- First row ---
    \begin{minipage}[b]{0.96\textwidth}
        \centering
        \includegraphics[width=\textwidth]{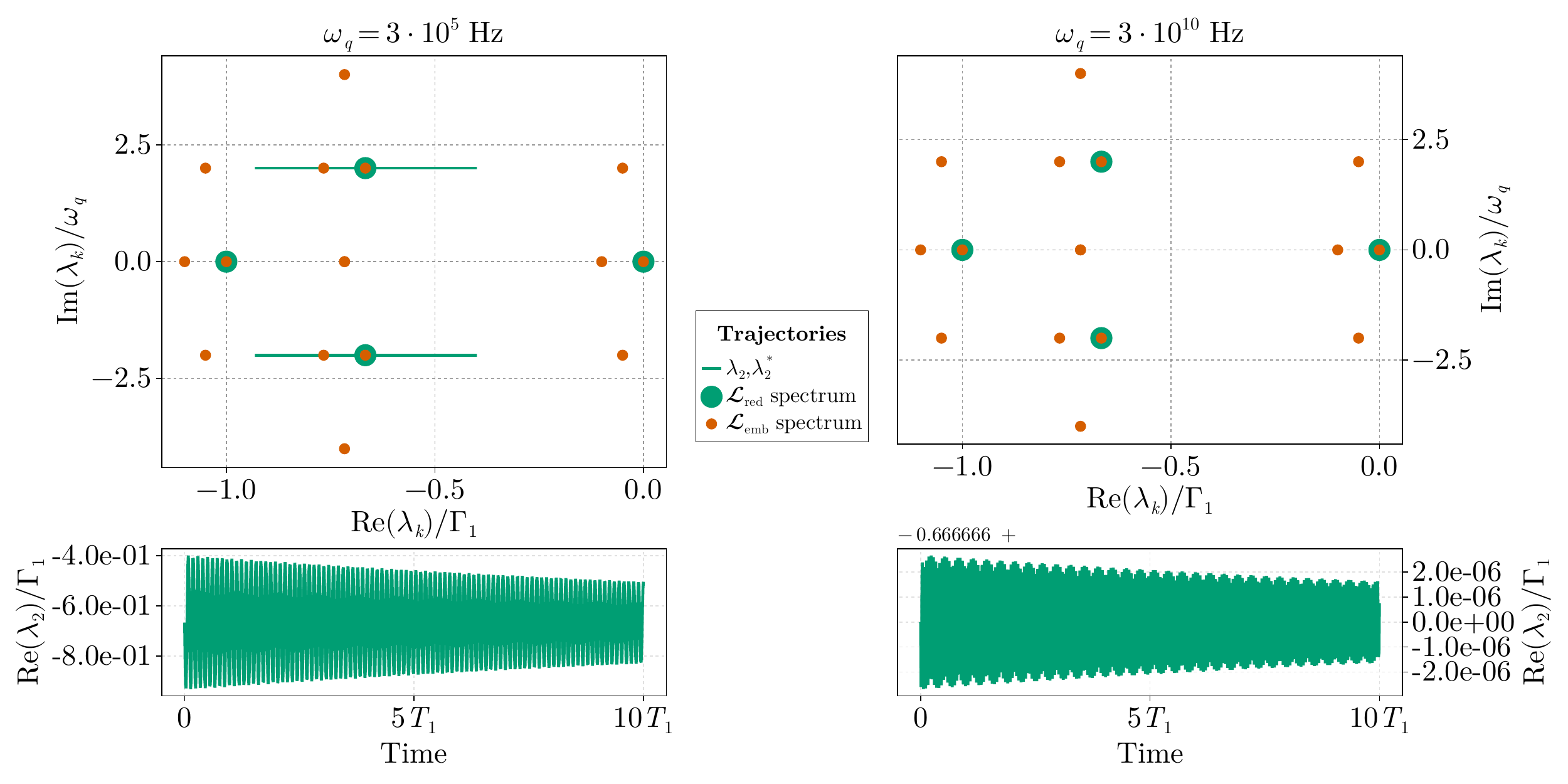}
        \textbf{(a)}
    \end{minipage}
    \vspace{0.5cm}
    % --- Second row ---
    \begin{minipage}[b]{0.55\textwidth}
        \centering
        \includegraphics[width=\textwidth]{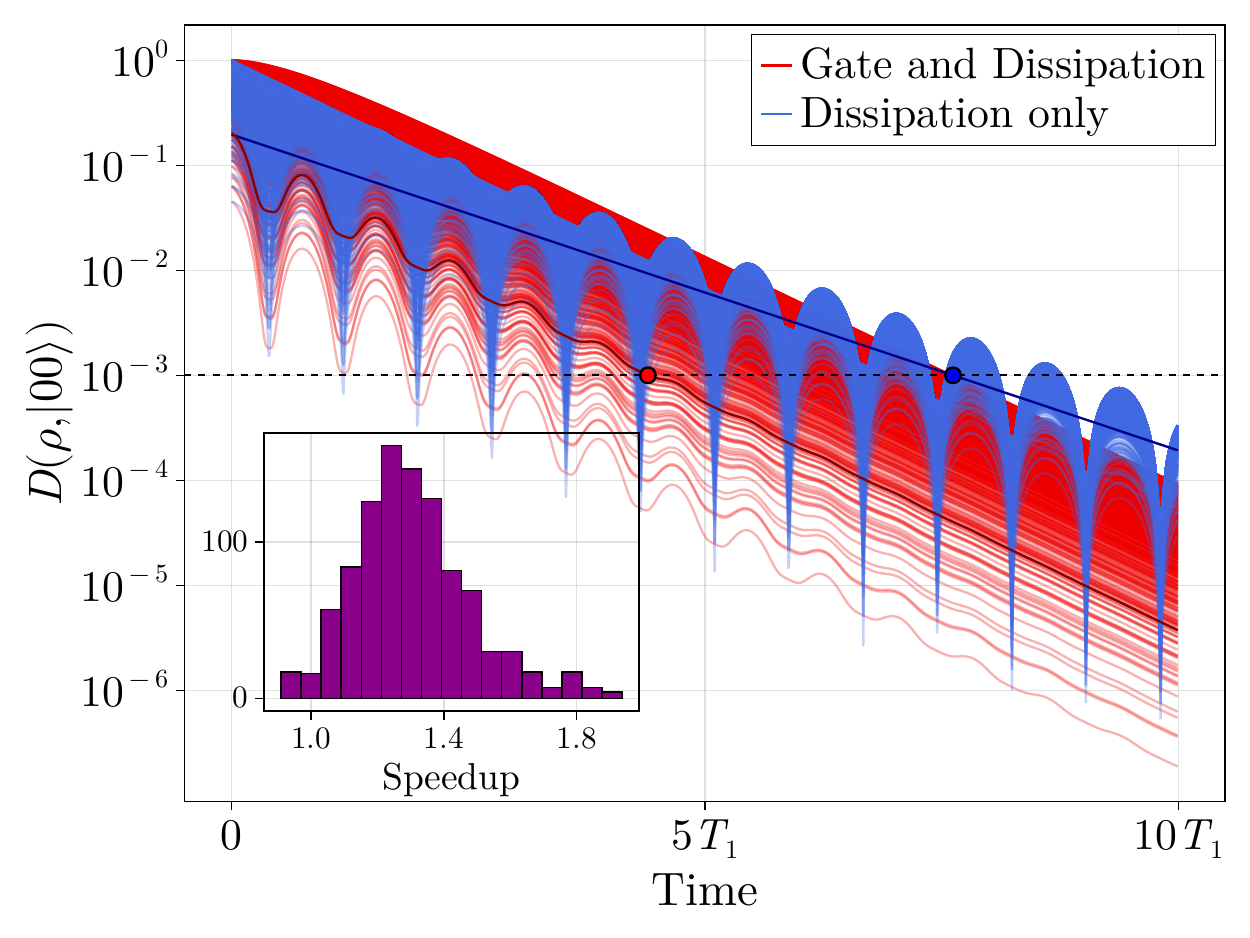}
        \textbf{(b)}
    \end{minipage}

    \caption{(a) Spectra of the full Markovian embedding [Eq.~(\ref{eq:non-Markovian:model})] (orange points) and of the reduced Liouvillian [Eq.~(\ref{eq:reduced_liouv})] as a function of time (green lines) and for long times (green points). The lower plots show the time evolution of $\text{Re}(\lambda_2)$ as a function of time, with $\omega_q = 3.10^{5}$ Hz (left plot) and $3.10^{10}$ Hz (right plot), while $\kappa/\nu_{zx} = 0.05$.
    (b) Trace distance to the ground state of $1000$ two-qubit states as a function of time, where the system qubit $q_1$ is sampled from the Haar measure and the ancilla and the TLSs are initialized in their ground state. Results are shown with (red) and without (blue) application of the C-Ry gate. The inset shows the histogram of achieved speedups for $\epsilon=10^{-3}$. While $\kappa/\nu_{zx}$ is set to 0.05, $\omega_q$ and $\omega_t$ are set to small values to explicitly display the non-Markovian behavior (i.e., the oscillations), even though the speedups achieved are very similar for larger frequencies. We perform exponential fits following the curves' peaks, such that the speedup corresponds to the relative reduction of the time it takes the whole curve to cross the horizontal line defined by $\epsilon$.
    }
    \label{fig:non-markovian:case}
\end{figure}
One physically motivated way to model non-Markovian noise for idling transmon qubits is to assume that each qubit is strongly coupled to a single two-level system defect (TLS), while the TLS itself interacts
with a Markovian environment~\cite{Burnett2019,Papic2022,Cho_2023,papic2023,Agarwal2024Modelling,Reiss2025}. 
\updated{Note that non-Markovianity can also be modeled using a Lindblad master equation with time-dependent rates that are occasionally negative, as was done, in the context of the Mpemba effect, in~\cite{Peluso_2026}.}
Although a transmon typically couples to
many defect TLSs, only those that lie close to resonance with the qubit contribute appreciably to its
dynamics, and in many situations, a single near-resonant TLS dominates the dissipative behavior over
the relevant timescales \cite{Cho_2023}.
In this description, the combined qubit-TLS system evolves under the Markovian
Liouvillian
\begin{equation}
\begin{aligned}
    &\mathcal{L}_{\mathrm{emb}}[\hat\rho] = \\& - i[\hat H,\hat\rho] + \Gamma_1 \left(\mathcal{D}_{\hat\sigma^{1}_-}\left[\hat\rho\right]+\mathcal{D}_{\hat\sigma^{2}_-}\left[\hat\rho\right]\right)
    + \frac{\Gamma_\phi}{2}\left( \mathcal{D}_{\hat\sigma^{1}_z}\left[\hat\rho\right]
    + \mathcal{D}_{\hat\sigma^{2}_z}\left[\hat\rho\right] \right)
    + \kappa \left(\mathcal{D}_{\hat\sigma^\mathrm{TLS_1}_-}\left[\hat\rho\right] + \mathcal{D}_{\hat\sigma^\mathrm{TLS_2}_-}\left[\hat\rho\right]\right),
    \end{aligned}
    \label{eq:non-Markovian:model}
\end{equation}
where $\hat H  = \omega_q(\hat\sigma^{1}_z + \hat\sigma^{2}_z) + \omega_t(\hat\sigma^{\mathrm{TLS}_1}_z + \hat\sigma^{\mathrm{TLS}_2}_z) + \nu_{zx} (\hat\sigma^{1}_z \otimes \hat\sigma^{\mathrm{TLS_1}}_x + \hat\sigma^{2}_z \otimes \hat\sigma^{\mathrm{TLS_2}}_x)$.
Each qubit therefore undergoes amplitude and phase damping as in the Markovian case
(see~\cref{eq:2qubit:davies:map:T0:with:dephasing}), while interacting with a TLS
of frequency $\omega_t$. \updated{In the standard tunneling model (STM), TLSs can be described using a double-well potential with an energy difference. In such model, a charge displacement between the two wells creates an electric dipole. The interaction with a qubit is modeled as a transverse-longitudinal one, and represents the coupling between the electric dipole of the TLS and the oscillating electric fields in capacitive circuit components and tunnel junction barriers with strength $\nu_{zx}$. The TLSs relaxation to their ground state at rate $\kappa$ can be mainly explained by the emission of acoustic phonons or photonic loss~\cite{Muller2019, Kumar2016}.} A sketch illustration of this model is displayed in \cref{fig:NM_sketch}.
It captures the behavior of a flux-tunable qubit coupled to a magnetic impurity, which is believed to be one of the main contributors to the pure dephasing in such circuits \cite{Kumar2016,Muller2019,Braumuller2020}. 
Tracing out the TLSs yields reduced qubit dynamics that become non-Markovian whenever $4\nu_{zx}^2 > \kappa^2/16$, as signaled by the appearance of damped purity oscillations~\cite{Papic2022, Cho_2023, Agarwal2024Modelling}.
In \Cref{app:non:markovian} we derive an effective Redfield generator $\mathcal{L}_{\mathrm{red}}$ for describing such dynamics. The resulting master equation for the physical qubit density matrix $\hat\rho_s$ reads\begin{equation}
    \begin{aligned}
        \frac{\mathrm{d} \hat{\rho}_s}{\mathrm{d}t} = \mathcal{L}_{\mathrm{red}}\left[ \hat{\rho}_s\right] = &-i\omega_q[\hat{\sigma}_z, \hat{\rho}_s] + \Gamma_1\mathcal{D}_{\hat{\sigma}_-}[\hat{\rho}_s] \\ &+\left(\frac{\Gamma_\phi}{2} + \frac{\nu^2_{zx}}{\frac{\kappa^2}{4}+4\omega^2_t}\left[ -\kappa e^{-\kappa t} + e^{\frac{-\kappa}{2} t} \left( \kappa \cos(2\omega_t t) + 4\omega_t\sin(2\omega_t t) \right)\right] \right)\mathcal{D}_{\hat{\sigma}_z}[\hat{\rho}_s] 
    \end{aligned}
    \label{eq:reduced_liouv}
\end{equation}
which displays a time-dependent phase damping rate, thus yielding eigenvalues that oscillate in time. Indeed, panel (a) of \cref{fig:non-markovian:case} displays the trajectories of the eigenvalues of the reduced Liouvillian $\mathcal{L}_{\mathrm{red}}$ over time (green lines) -- along with their values at long times (green dots) -- and the spectrum of the full Markovian embedding Liouvillian $\mathcal{L}_{\mathrm{emb}}$ (orange dots), while the lower plots display $\text{Re}(\lambda_2)$ as a function of time. We used parameter values taken from~\cite{Agarwal2024Modelling} that have been shown to fit well transmon qubit experimental data. The spectrum of $\mathcal{L}_\mathrm{red}$ perfectly corresponds to a subset of the full description, whether the qubit frequency is small (left) or large (right), displaying the wide validity regime of the effective reduced model. 
\begin{figure}[!h]
    \centering
    \begin{minipage}[b]{0.49\textwidth}
        \centering
        \includegraphics[width=\textwidth]{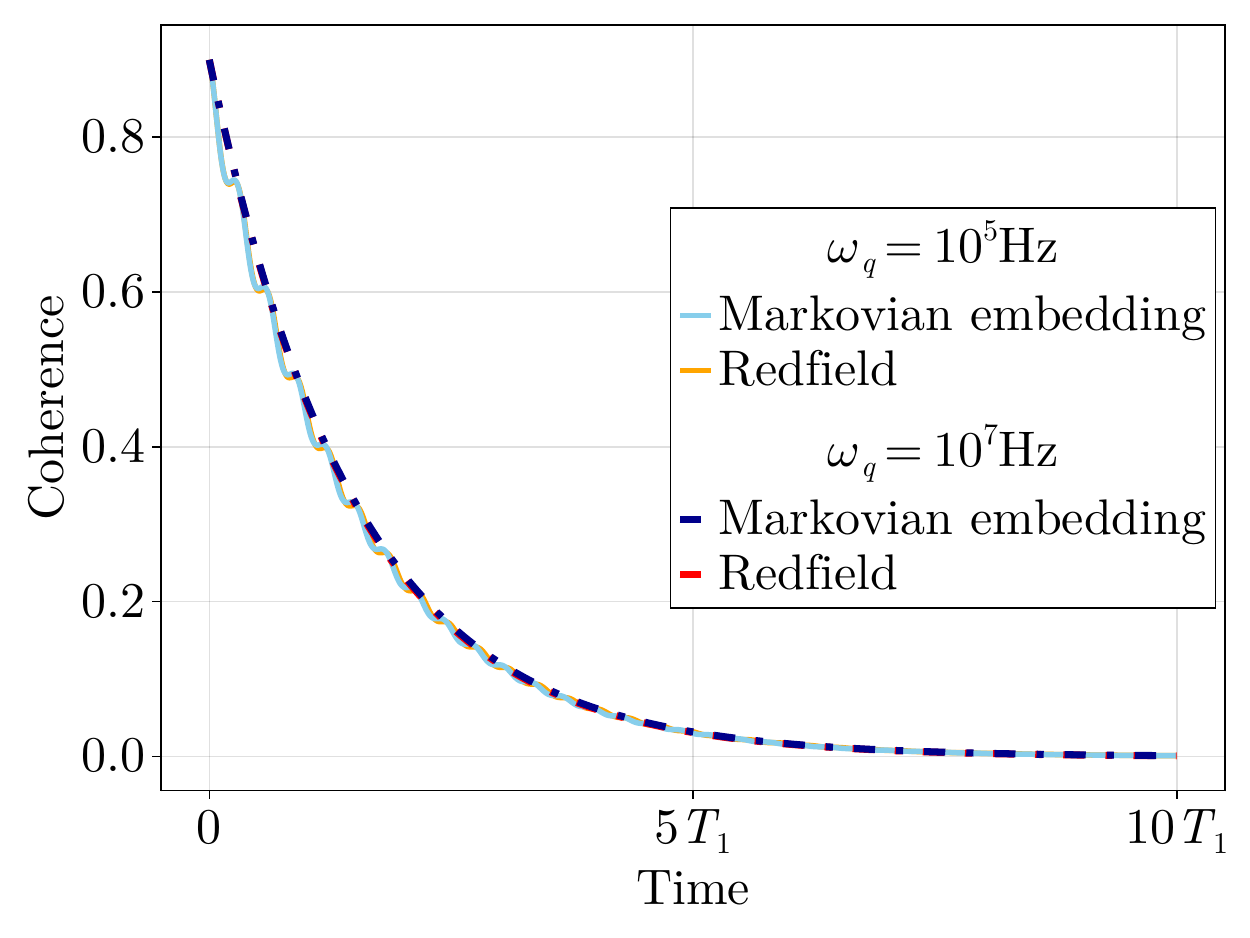}
        \textbf{(a)}
    \end{minipage}
    \hfill
    \begin{minipage}[b]{0.49\textwidth}
        \centering
        \includegraphics[width=\textwidth]{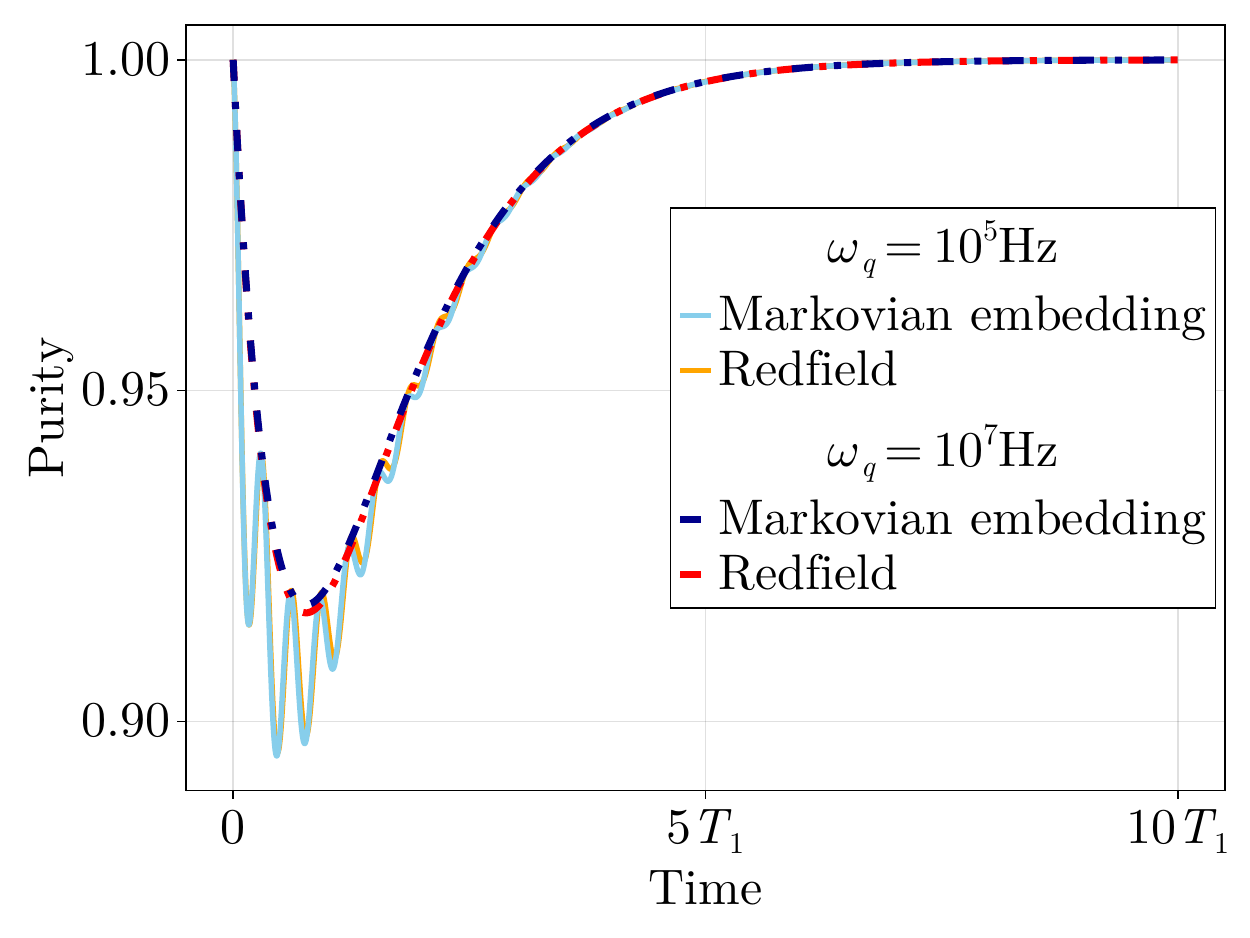}
        \textbf{(b)}
    \end{minipage}
    \caption{Comparison of the dynamics of the two-qubit state coherence [panel (a)] and purity [panel (b)], obtained from the Markovian embedding of \cref{eq:non-Markovian:model} (blue curves) and the reduced Redfield model of \cref{eq:reduced_liouv} (orange curves). Both are evaluated for $\omega_q = 10^5$ Hz (solid line) and $\omega_q = 10^7$ Hz (dashed line), while $\kappa/\nu_{zx} = 0.05$.
    }\label{fig:emb_vs_redfield_pur_coh}
\end{figure}
Unlike the Markovian case, here the spectral ratio obtained from $\mathcal{L}_{\mathrm{red}}$ defines a time-dependent speedup that reads
\begin{equation}
\begin{aligned}
S_\mathrm{red}(t)&=|\text{Re}(\lambda_3(t))|/|\text{Re}(\lambda_2(t))| = \\ &\frac{|2\Gamma_1(\kappa^2+16\omega_t^2)|}{|-(\Gamma_1+2\Gamma_\phi)(\kappa^2+16\omega_t^2)+16\nu^2_{zx}e^{-\kappa t/2}\left(\kappa(e^{-\kappa t/2}-\cos(2\omega_t t))-4\omega_t\sin(2\omega_t t)\right)|}.    
\end{aligned}
\end{equation}
For long times, this tends to 
\begin{equation}
    S_{\mathrm{red}} = \frac{|2\Gamma_1|}{|-(\Gamma_1+2\Gamma_\phi)|} = \frac{T_2}{T_1},\label{eq:speedup:redfield}
\end{equation}
which corresponds to the Markovian case [Eq.~(\ref{eq:asymptotic_speedup})].
On the other hand, the full Markovian embedding obtained from $\mathcal{L}_{\mathrm{emb}}$ yields a speedup
\begin{equation}
S_{\mathrm{emb}} \approx \frac{|2\Gamma_1|}{|\frac{ \kappa^3 \nu_{zx}^4}{8\omega_t^6} - \frac{\kappa \nu_{zx}^2 (\kappa^2 + 16 \nu_{zx}^2)}{32\omega_t^4} + \frac{\kappa \nu_{zx}^2}{2\omega_t^2} - (\Gamma_1 + 2 \Gamma_\phi + \kappa)|},
\label{eq:emb:speedup}
\end{equation}
after performing a fourth-order series expansion in the small parameters $\xi_1 = \kappa/\omega_t$ and $\xi_2=\nu_{zx}/\omega_t$.
In the regime $\xi_1, \xi_2 \rightarrow 0$, \cref{eq:emb:speedup} becomes
\begin{equation}
    S_{\mathrm{emb}} \approx  \frac{|2\Gamma_1|}{|-(\Gamma_1 + 2 \Gamma_\phi + \kappa)|} = \frac{T_2/T_1}{1+\kappa T_2/2}.
    \label{eq:emb:speedup:simpl}
\end{equation}
This correction term in $\kappa$ is due to the qubit-TLS coupling acting as dephasing on the qubit, thus decreasing its effective coherence time.

Importantly, these simplified expressions are obtained in the long-time limit, while for the reset protocols considered here transient dynamics play a central role.
To demonstrate the applicability of our approach to arbitrary coherent states, we apply it, similarly to the Markovian case, to random pure states for the qubit $q_1$, while the ancilla qubit $q_2$ and the TLSs (one for each physical qubit) are initialized in their ground state. The trace distances to the ground state of the two-qubit state, i.e., the partial trace over the two TLSs, with and without application of a C-Ry gate, as a function of time, are displayed in panel (b) of \cref{fig:non-markovian:case} in blue and orange, respectively. The panel's inset shows the achieved speedup distribution, with a median speedup around $1.3$. This follows from \cref{eq:emb:speedup:simpl}, which yields a speedup of 
$1.39$ ($T_2/T_1 = 1.5$), multiplied by a finite $\epsilon$ correction at short times.
Finally, we compare the Markovian embedding and Redfield approach by analyzing the two-qubit state coherence and purity as a function of time, for different values of $\omega_q$. The results are displayed in \cref{fig:emb_vs_redfield_pur_coh}, showing an excellent fit of the Redfield curves in both parameter regimes.
In conclusion, even though~\cref{eq:speedup:redfield,eq:emb:speedup:simpl} show that the reduced model $\mathcal{L}_\mathrm{red}$ does not perfectly predict the speedup, it accurately reproduces the evolution of observables. Moreover, it allows for the identification of the qubit-associated eigenvalues in the Markovian embedding Liouvillian spectrum, which are essential for the speedup $S_{\mathrm{emb}}$ computation.
\section{Robustness under imperfect control}
\label{sec:imperfect:control}
Any practical implementation of our protocol must account for the fact that quantum gates are never
executed perfectly. Since the entangling C-Ry operation is responsible for suppressing the control
qubit’s coherences, coherent control errors in this gate can directly affect the achievable speedup.
\begin{figure}[h]
    \centering
    % --- First row ---
    \begin{minipage}[b]{0.32\textwidth}
        \centering
        \includegraphics[width=\textwidth]{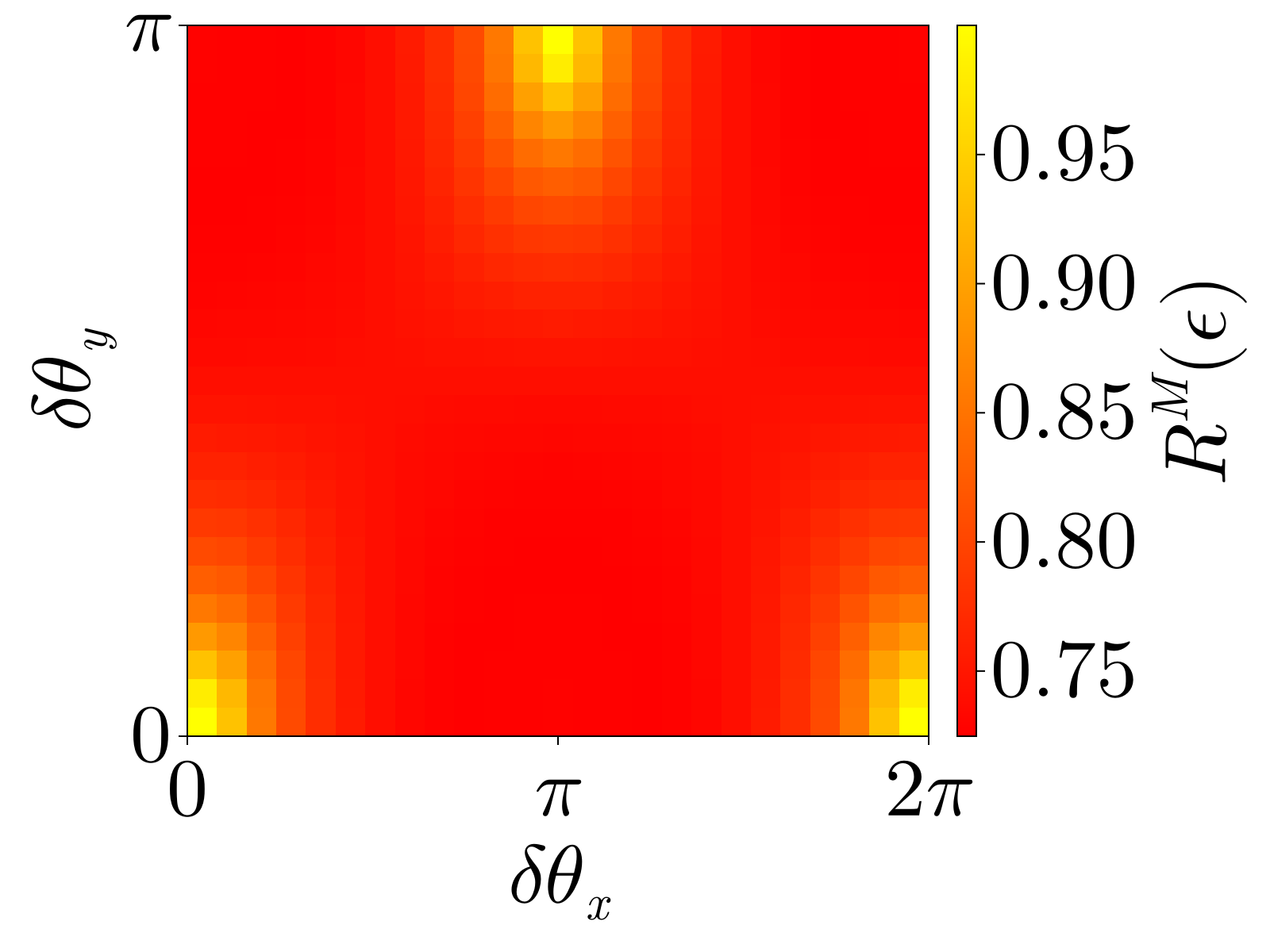}
        \textbf{(a)}
    \end{minipage}
    \hfill
    \begin{minipage}[b]{0.32\textwidth}
        \centering
        \includegraphics[width=\textwidth]{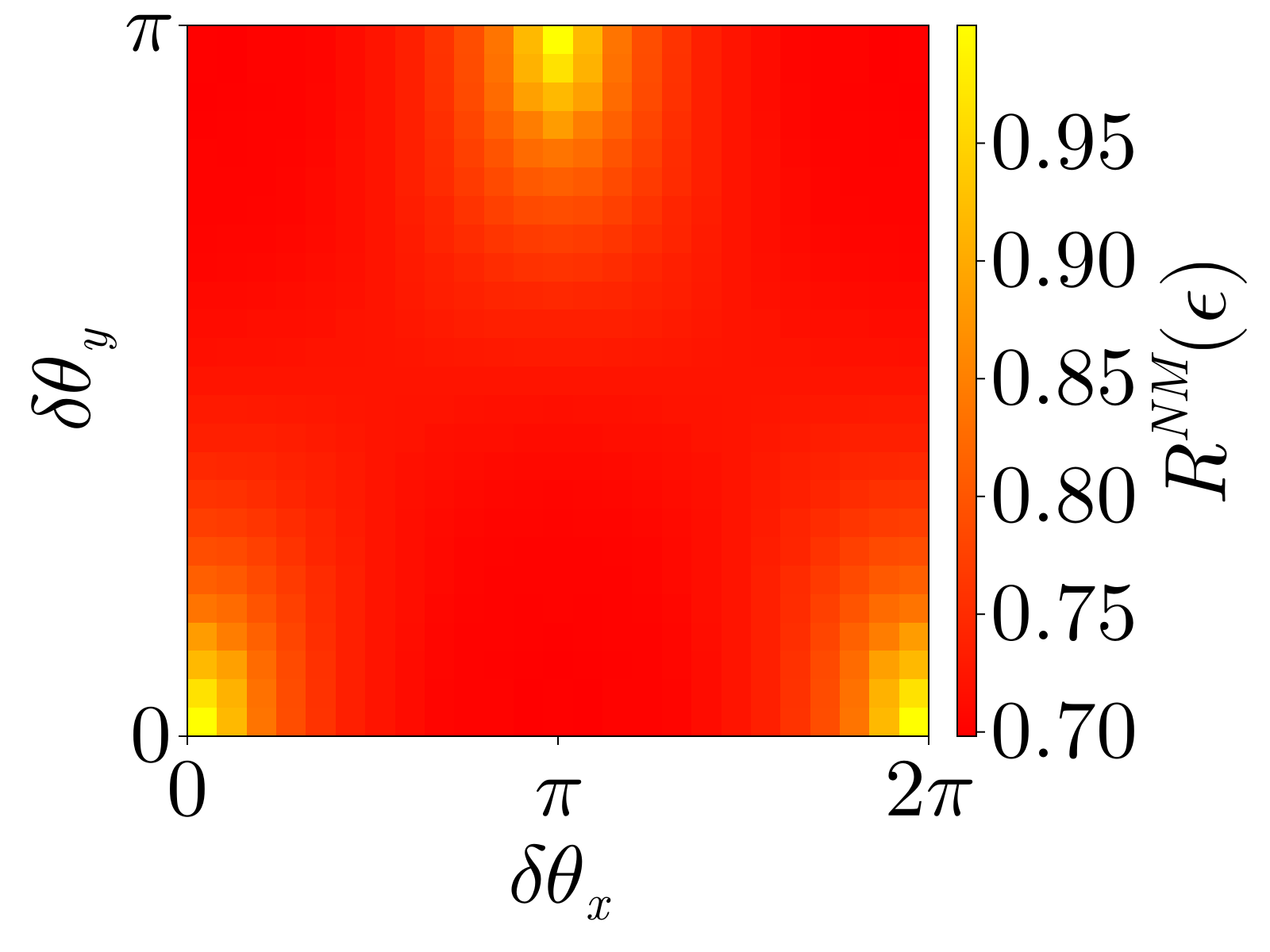}
        \textbf{(b)}
    \end{minipage}
    \hfill
    \begin{minipage}[b]{0.32\textwidth}
        \centering
        \includegraphics[width=\textwidth]{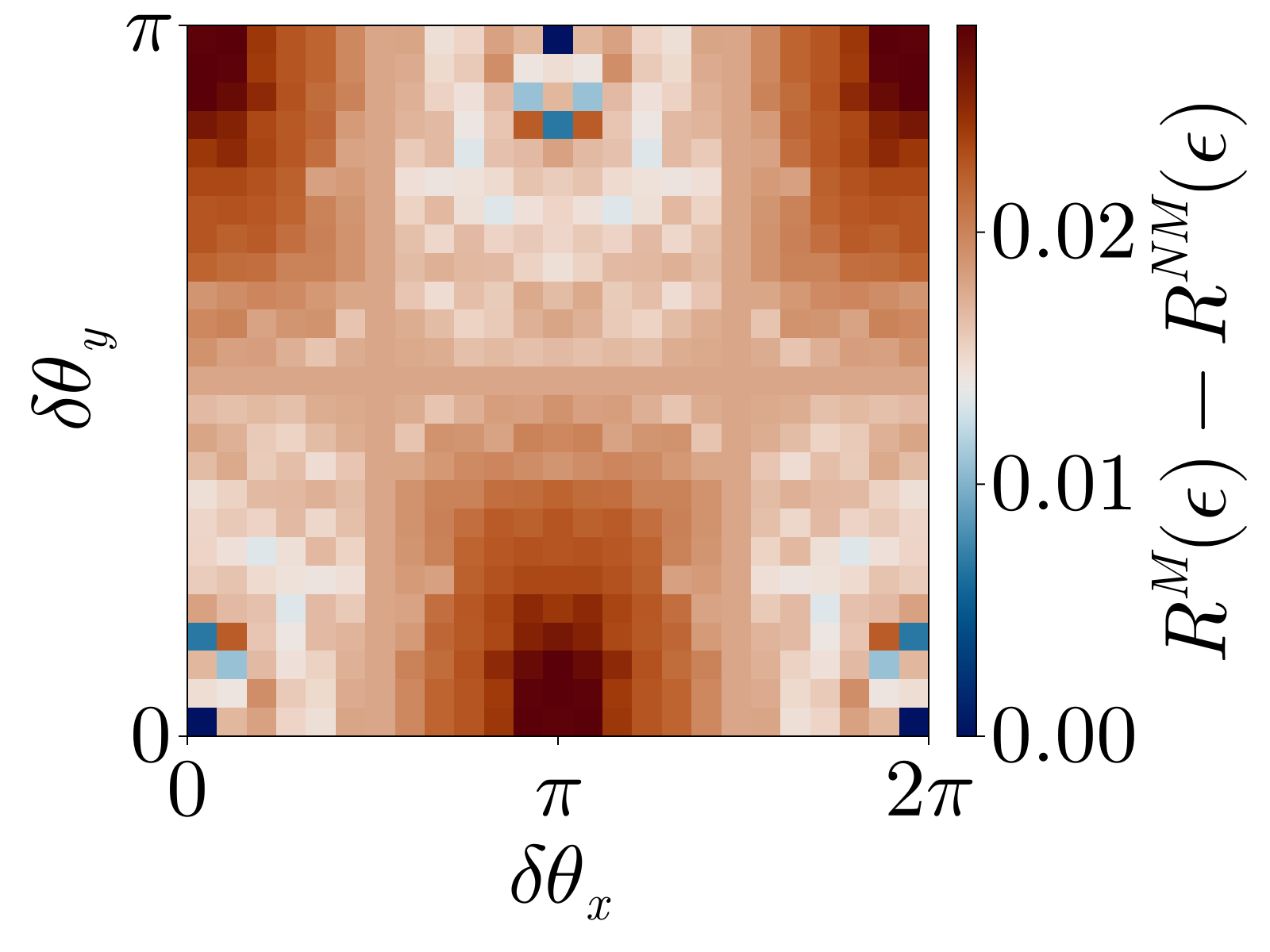}
        \textbf{(c)}
    \end{minipage}
    \caption{Robustness of the C-Ry–based reset protocol under imperfect control. The robustness measure $R(\epsilon)$~\cref{eq:imperfect:control} is shown as a function of systematic rotation errors $\delta\theta_y$ in the intended conditional $y$ rotation and spurious rotations $\delta\theta_x$ about the $x$ axis. Panel (a) considers the Markovian model~\cref{eq:2qubit:davies:map:T0:with:dephasing}, panel (b) corresponds to the non-Markovian model~\cref{eq:non-Markovian:model} and panel (c) shows the difference between the two. In both panels, $R(\epsilon)=1$ indicates the relaxation time obtained with an ideal C-Ry gate, while deviations quantify the impact of coherent control imperfections. The robustness has been averaged over $50$ Haar random states for $\hat\rho_1$, while we fixed $\hat\rho_2 = \ket{0}\!\bra{0}$.
    }
    \label{fig:imperfect_control}
\end{figure}
We model \textit{imperfect control} as systematic under- or over-rotations of the target qubit. Such miscalibrations may occur both in the intended $y$ rotation and through
spurious rotations around the $x$ axis, leaving residual coherences that limit the effectiveness of the
protocol. It is therefore important to quantify how sensitive the accelerated reset is to these two types
of coherent errors.
To this end, we compare the relaxation time achieved with an imperfectly implemented C-Ry gate
to that obtained under ideal control. We define a measure for the \textit{robustness} of the reset
speedup against coherent control imperfections as
\begin{equation}
    R(\epsilon) = \frac
{t(\epsilon, \hat\rho_{\mathrm{ex}})}{t(\epsilon, \hat\rho_{\mathrm{err}})}\,,
    \label{eq:imperfect:control}
\end{equation}
where $\hat\rho_{\mathrm{ex}}$ and $\hat\rho_{\mathrm{err}}$ are the two-qubit states obtained after the exact and imperfect gate applications, respectively.
The protocol effectively accelerates qubits reset when no errors occur in the gate, as shown by the yellow areas in the lower corners of~\cref{fig:imperfect_control}. However, this speedup also occurs, with a similar scale, for rotation errors of $\pi$ along the $x$ and $y$ axis, in which case the total $y$-rotation angle is $2\pi$, thus equivalent to applying the identity, and the C-Rx gate acts as a dephasing channel on the control qubit.
Interestingly,~\cref{fig:imperfect_control} also shows that the robustness of the protocol is very similar for the Markovian model [c.f.~\cref{eq:2qubit:davies:map:T0:with:dephasing}, panel (a)] and the non-Markovian one [c.f.~\cref{eq:non-Markovian:model}, panel (b)], with the latter one being slightly smaller, as shown in panel (c).

\section{Experimental Implementation} 
\label{sec:experiment}
In this section, we experimentally demonstrate the suppression of local single-qubit coherences by
redistributing them into global two-qubit coherences via the application of an entangling gate.
The experiment is performed on the IQM Garnet superconducting quantum processor, accessed via
cloud services \cite{IQM2024}. While our theoretical analysis focused on a controlled-$R_y$ gate, here
we employ a CNOT gate, which produces the same delocalization of coherences and therefore realizes
the same Mpemba-enhanced reset mechanism (see~\cref{sec:cnot:protocol}).

To reconstruct the effective qubit dynamics, we perform three characterization experiments:
\begin{enumerate}
    \item \textit{$T_1$ measurement:} a $\pi$ pulse is applied to the qubit, followed by a waiting time
    $\Delta t$, after which the excited-state population is measured as a function of $\Delta t$.
    \item \textit{$T_2$ measurement with dynamical decoupling:} a $\pi/2$ pulse is applied, followed by
    a waiting time $\Delta t$, a $-\pi/2$ pulse, and a measurement of the excited-state population as a
    function of $\Delta t$.
    \item \textit{$T_2$+CNOT measurement with dynamical decoupling:} identical to the protocol in point~2, except that a CNOT
    gate is applied during the waiting period, as described in Sec.~\ref{sec:cnot:protocol}. The CNOT
    is implemented using a native CZ gate and additional single-qubit rotations.
\end{enumerate}
Dynamical decoupling is employed to suppress low-frequency noise contributions to pure dephasing,
allowing us to access the regime $T_2 > T_1$ \cite{Cywinski2008,papic2023}. For all three experiments,
we fit exponential decay curves and neglect the duration of the single-qubit pulses (on the order of
10~ns) \cite{IQM2024}. The $T_1$ measurement determines the decay of the diagonal elements of the
density matrix, while the latter two measurements probe the decay of the off-diagonal elements.

\begin{figure}[h]
    \centering
    \includegraphics[width=0.5\linewidth]{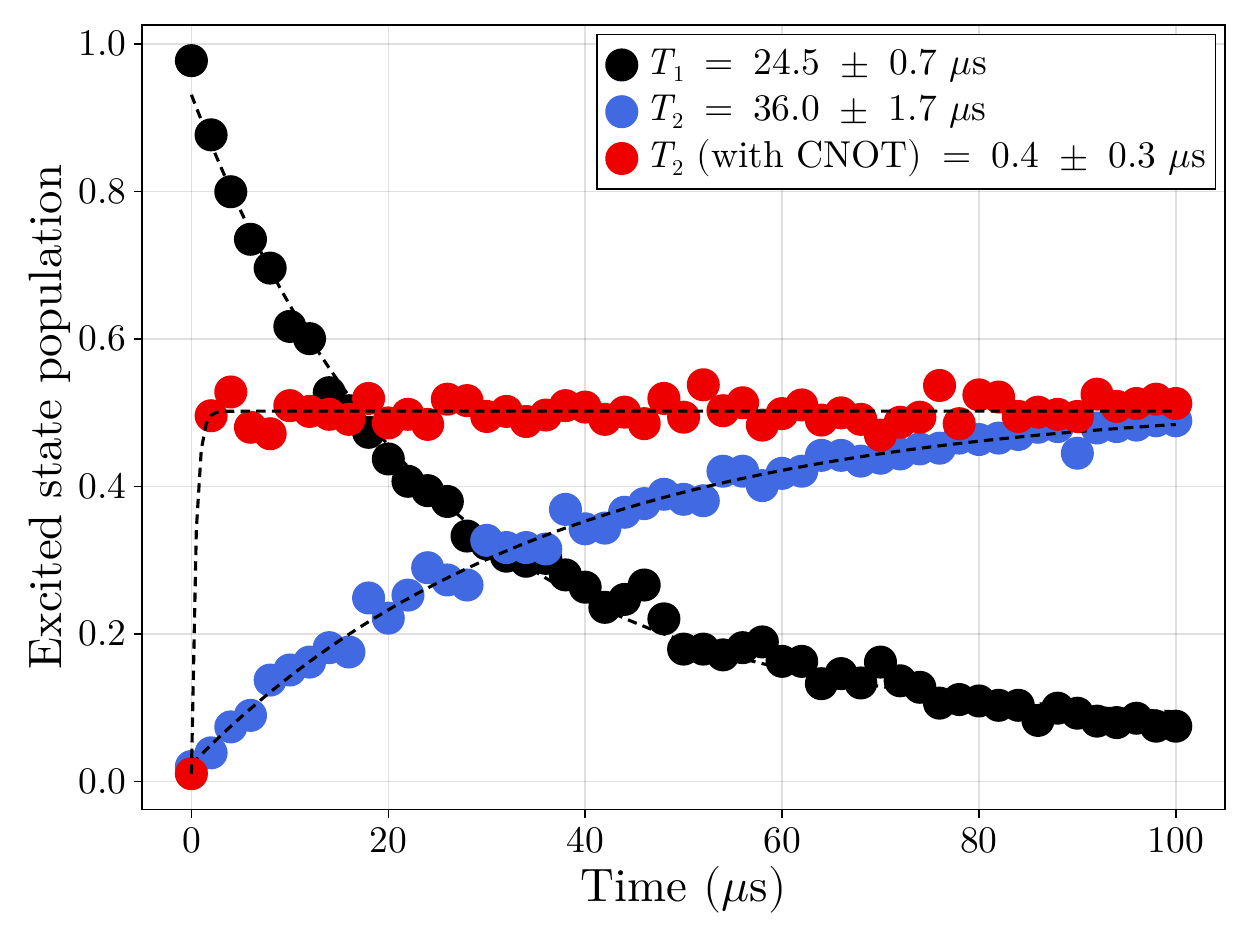}
    \caption{Decay time measurements from IQM Garnet, illustrating the rapid suppression of local qubit coherences via an entangling gate. Assuming near-ideal single-qubit gate and measurement
    fidelity, the excited-state probability corresponds to $\rho_{11}(\Delta t)$
    for the $T_1$ measurement (black) and to $\frac{1}{2}\!\left(1 - |\rho_{01}
    (\Delta t)|\right)$ for both $T_2$ measurements (red and blue). In the latter case, the value $1/2$
    corresponds to an incoherent state. Each data point
    is obtained from 1024 shots, and all measurements are performed sequentially to minimize temporal
    drifts.}
    \label{fig:coherence:measurements}
\end{figure}
The results in \cref{fig:coherence:measurements} show that the application of the CNOT gate strongly
suppresses the local coherence of the target qubit on a timescale much shorter than both the intrinsic
$T_1$ and $T_2$ decay times. As discussed in Sec.~\ref{sec:cnot:protocol}, this arises because the CNOT gate converts local $q_1$ coherences into global two-qubit coherences, which decay at an enhanced rate
under dissipation.

Taken together, these measurements allow us to reconstruct the evolution of an arbitrary initial qubit
state under Markovian $T_1$ and $T_2$ dynamics, both with and without the Mpemba-enhanced reset
protocol, according to
\begin{equation}
    \hat{\rho}(t) =
    \begin{pmatrix}
        \rho_{00}(0) + \rho_{11}(0)\!\left(1 - e^{-t/T_1}\right)
        & \rho_{01}(0)\, e^{-t/T_2} \\
        \rho_{01}^*(0)\, e^{-t/T_2}
        & \rho_{11}(0)\, e^{-t/T_1}
    \end{pmatrix}.
\end{equation}
In the present experiment, this yields the asymptotic speedup
$\lim_{\epsilon\to 0} S(\epsilon) = T_2/T_1 \approx 1.47$.

\section{Conclusion and Outlook}
\label{sec:conclusion:outlook}
In this work, we have demonstrated how the quantum Mpemba effect can be harnessed to accelerate passive qubit reset.
By exploiting the fact that states with vanishing overlap with the slowest decaying Liouvillian mode thermalize exceptionally fast, we showed that it is possible to substantially reduce reset times without resorting to measurement, feedback, or engineered dissipation.
Our protocol relies on a single entangling operation between the target qubit and a neighboring post-measurement qubit, making it conceptually simple and compatible with existing superconducting-qubit architectures.
We focused on the experimentally relevant regime $T_2 > T_1$, where long-lived coherences dominate the late-time relaxation dynamics and demonstrated that passive reset can be accelerated by delocalizing single-qubit coherences: A single entangling gate (e.g., CNOT or CRY$(\pi)$) between a qubit in an arbitrary state and an incoherent ancilla converts local coherences into fast-decaying global two-qubit coherences, thereby eliminating the overlap with the slowest-decaying Liouvillian mode and yielding a speedup of approximately $T_2/T_1$. In quantum algorithms where only part of the register is measured~\cite{IBM2023, Google2025Echoes}, the resulting incoherent qubits provide a readily available resource to accelerate the reset of the remaining qubits.

Beyond the idealized Markovian setting, we extensively analyzed the robustness of the protocol under more realistic conditions.
We showed that the accelerated reset persists in the presence of non-Markovian noise induced by coupling to environmental two-level systems, as well as at finite temperature. 
Furthermore, we quantified the impact of imperfect coherent control and demonstrated that the protocol remains effective over a broad range of systematic gate errors.
These conclusions are reinforced by our experimental results obtained on a superconducting quantum processor.
Taken together, these indicate that Mpemba-enhanced passive reset is remarkably resilient to experimentally relevant imperfections.

Looking ahead, several interesting directions emerge from our work.
A natural extension would be to explore whether similar ideas can be applied in the opposite regime $T_1 > T_2$.
While a straightforward generalization is not possible without partial information about the qubit's state, identifying scenarios in which such knowledge is available (or can be inferred) could open new possibilities for accelerated reset.
Furthermore, from a thermodynamic perspective, qubit reset can be viewed as an information-erasure process in the sense of Landauer~\cite{Landauer1961,Parrondo2015,Miller2020}, suggesting interesting future directions for connecting accelerated relaxation to thermodynamic costs.
More broadly, our results highlight how nonequilibrium relaxation phenomena, such as the Mpemba effect, can be repurposed as practical tools in quantum information processing. 
We expect that similar strategies could be employed in other contexts where fast equilibration or state preparation is required, including dissipative quantum state engineering~\cite{Zhan2025,Lin2025, Lloyd2025PRXQuantum, Mi2024, Langbehn2025} and algorithmic cooling protocols~\cite{Schulman1999, Boykin2002, Raeisi2015, Xuereb2025}. 
Exploring these connections further is an exciting direction for future work.

\ack{We thank Alessandro Summer and Laetitia Bettmann for enlightening discussions. TL acknowledges support from the French Community of Belgium in the form of a FRIA doctoral scholarship.  F.C.B. acknowledges support from Taighde Éireann - Research Ireland under grant number IRCLA/2022/3922. JG and MM acknowledge funding from the Royal Society and Research Ireland. 
}

\section*{Code and Data Availability}
All code and data used to generate the figures in this work are available \href{https://github.com/MattiaMoroder/fast_qubit_reset_with_Mpemba.git}{here}.

\newpage

\appendix
\FloatBarrier

\section{The impact of finite temperature}
\label{app:finite:temperature}
In this appendix, we investigate the impact of finite temperature in the Markovian model
\cref{eq:2qubit:davies:map:T0:with:dephasing} considered in the main text.
For two non-interacting qubits of identical frequency $\omega$, the finite-temperature
Davies master equation with additional pure local dephasing reads
\begin{equation}
\mathcal{L}\left[\hat\rho\right]
= -i[\hat{H},\hat{\rho}]
+ \sum_{i=1}^2 \left(
\Gamma_{\downarrow}\,\mathcal{D}_{\hat{\sigma}_-^{i}}\left[\hat{\rho}\right]
+ \Gamma_{\uparrow}\,\mathcal{D}_{\hat{\sigma}_+^{i}}\left[\hat{\rho}\right]
+ \frac{\Gamma_{\phi}}{2}\,\mathcal{D}_{\hat{\sigma}_z^{i}}\left[\hat{\rho}\right]
\right).
\label{eq:finite:temperature:davies:map:with:dephasing}
\end{equation}
The excitation and relaxation rates satisfy the detailed-balance condition
$\Gamma_{\uparrow}/\Gamma_{\downarrow} = e^{-\beta \omega}$,
with $\beta = (k_B T)^{-1}$.
For a bosonic thermal bath, one has
$\Gamma_{\downarrow} = \gamma \big(\bar{n}(\omega)+1\big)$ and
$\Gamma_{\uparrow} = \gamma\,\bar{n}(\omega)$,
where $\bar{n}(\omega) = (e^{\beta\omega}-1)^{-1}$.
Note that we have assumed the pure dephasing noise to be temperature-independent.
The longitudinal relaxation time $T_1$ is given by
\begin{equation}
\frac{1}{T_1}
= \Gamma_{\downarrow} + \Gamma_{\uparrow}
= \gamma \big(2\bar{n}(\omega) + 1\big),
\end{equation}
and the steady-state excited-state population reads
$p_{e,\mathrm{ss}} = \Gamma_{\uparrow}/(\Gamma_{\downarrow}+\Gamma_{\uparrow})
= (1+e^{\beta\omega})^{-1}$.
The transverse coherence decay time $T_2$ is
\begin{equation}
\frac{1}{T_2}
= \frac{\Gamma_{\downarrow} + \Gamma_{\uparrow}}{2} + \Gamma_{\phi}
= \frac{1}{2T_1} + \Gamma_{\phi}.
\end{equation}

Crucially, the Lindbladian~\cref{eq:finite:temperature:davies:map:with:dephasing}
retains the same block-diagonal structure as~\cref{eq:2qubit:davies:map:T0:with:dephasing},
with one block governing the evolution of the qubit's populations and another governing the evolution of their coherences.
Consequently, in the regime $T_2 > T_1$, the application of an entangling gate (such as $CR_y(\pi)$ or CNOT) accelerates the qubit reset also at finite temperature.
Moreover, in analogy to~\cref{eq:l2_l3}, we have that the asymptotic speedup at finite temperature is given by
\begin{equation}
\label{eq:T2_over_T1_finiteT}
S(\epsilon\to 0) =\frac{T_2}{T_1}
= \frac{2\big(2\bar{n}(\omega)+1\big)}
{\big(2\bar{n}(\omega)+1\big)+2\Gamma_{\phi}/\gamma} \leq 2\,.
\end{equation}
Note that $T_2/T_1$ increases upon increasing the temperature and that the bound is saturated at any temperature for $\Gamma_\phi=0$. 

\section{Robustness of the protocol with respect to different ancilla states} \label{app:robustness:ancilla}
\begin{figure}
    \centering
    \includegraphics[width=0.5\linewidth]{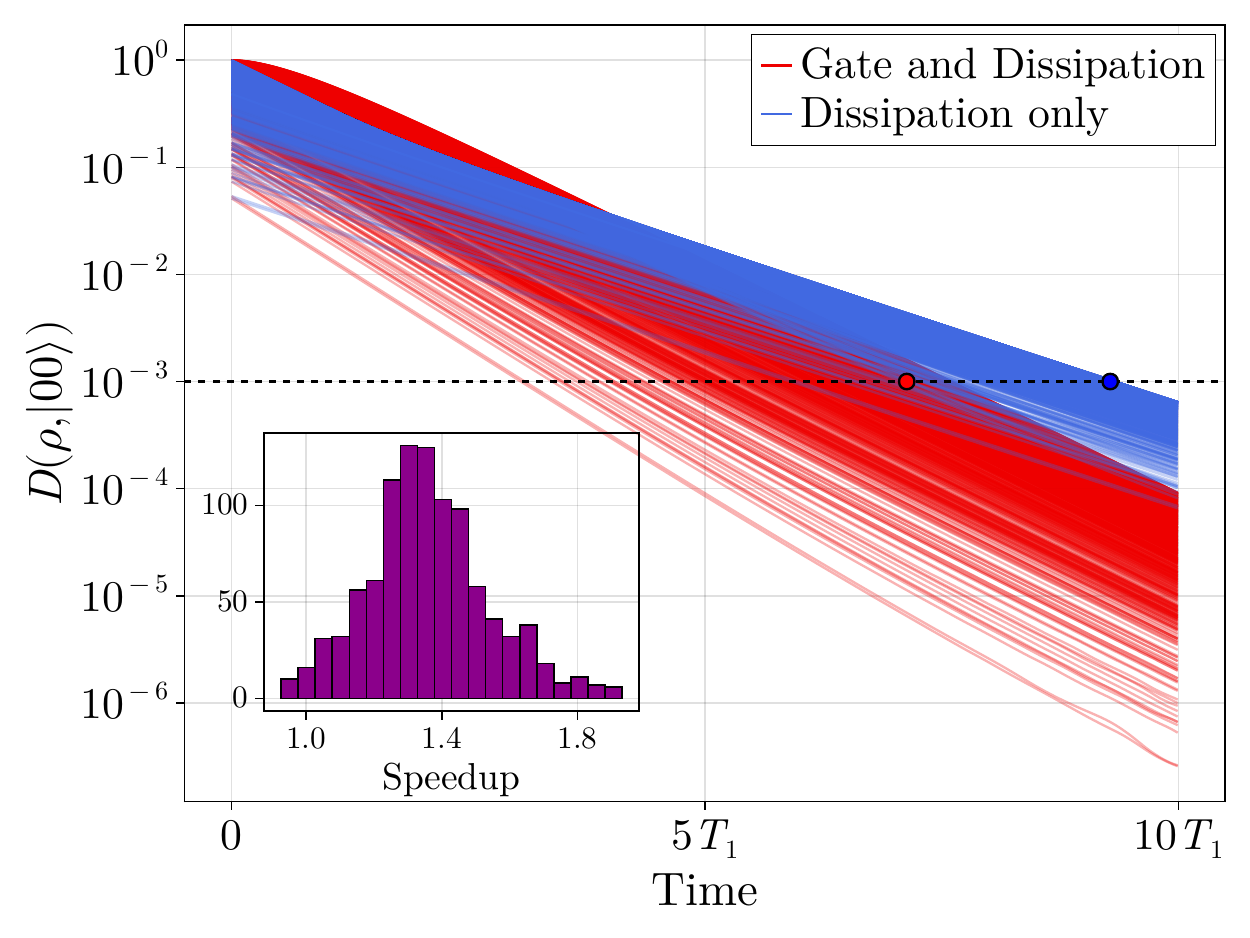}
    \caption{Trace distance to the ground state as a function of time (main plot) and corresponding speedup histogram for $\epsilon=10^{-3}$ (inset), obtained using the same parameters as in panel~(d) of~\cref{fig:markovian}, but with the ancilla qubit initialized in the ground state $\ket{0}$ rather than in the excited state $\ket{1}$.
    }
    \label{fig:app:speedup:q2:zero}
\end{figure}
In this appendix, we analyze how the achievable reset speedups depend on the initial state of the ancilla qubit.
Throughout, we focus on the Markovian model introduced in~\cref{sec:markovian:case}.
As discussed in the main text, the controlled $R_y(\pi)$ gate acts as an ideal dephasing channel on the system qubit $q_1$ only when the ancilla qubit $q_2$ is initially in an incoherent state, i.e., diagonal in the computational basis.
In panel~(d) of~\cref{fig:markovian}, we presented the distribution of speedups obtained for Haar-random initial states of $q_1$ when the ancilla is initialized in the excited state $\ket{1}$.
Here, we consider the complementary case where the ancilla is initialized in the ground state $\ket{0}$.
\Cref{fig:app:speedup:q2:zero} shows the trace distance to the ground state as a function of time for this choice of ancilla initialization, together with the corresponding histogram of speedups.
We find a median speedup of $\approx 1.4$, which is close to the results obtained for an ancilla prepared in $\ket{1}$.

\begin{figure}[h]
    \centering
    % --- First row ---
    \begin{minipage}[b]{0.48\textwidth}
        \centering
        \includegraphics[width=\textwidth]{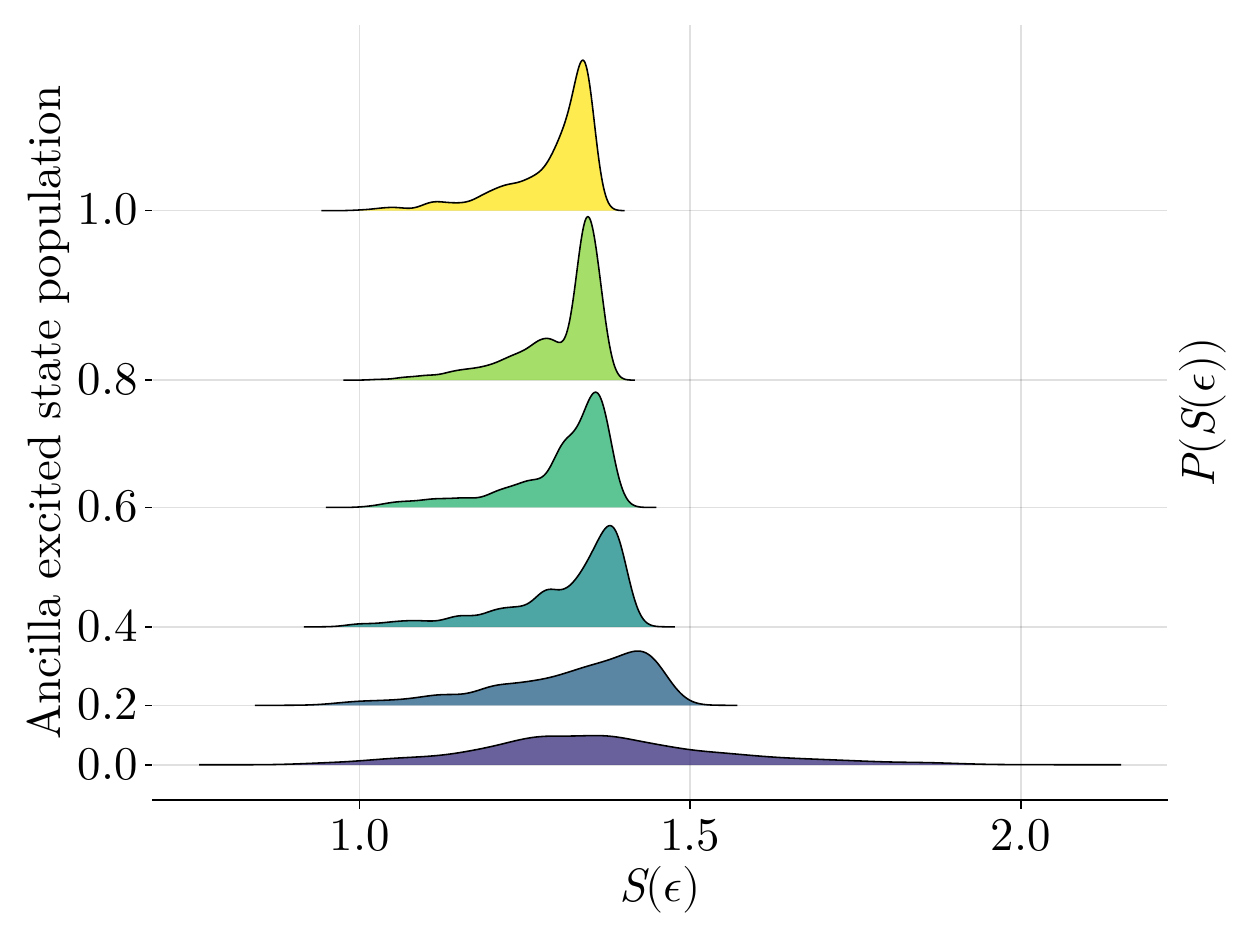}
        \textbf{(a)}
    \end{minipage}
    \hfill
    \begin{minipage}[b]{0.48\textwidth}
        \centering
        \includegraphics[width=\textwidth]{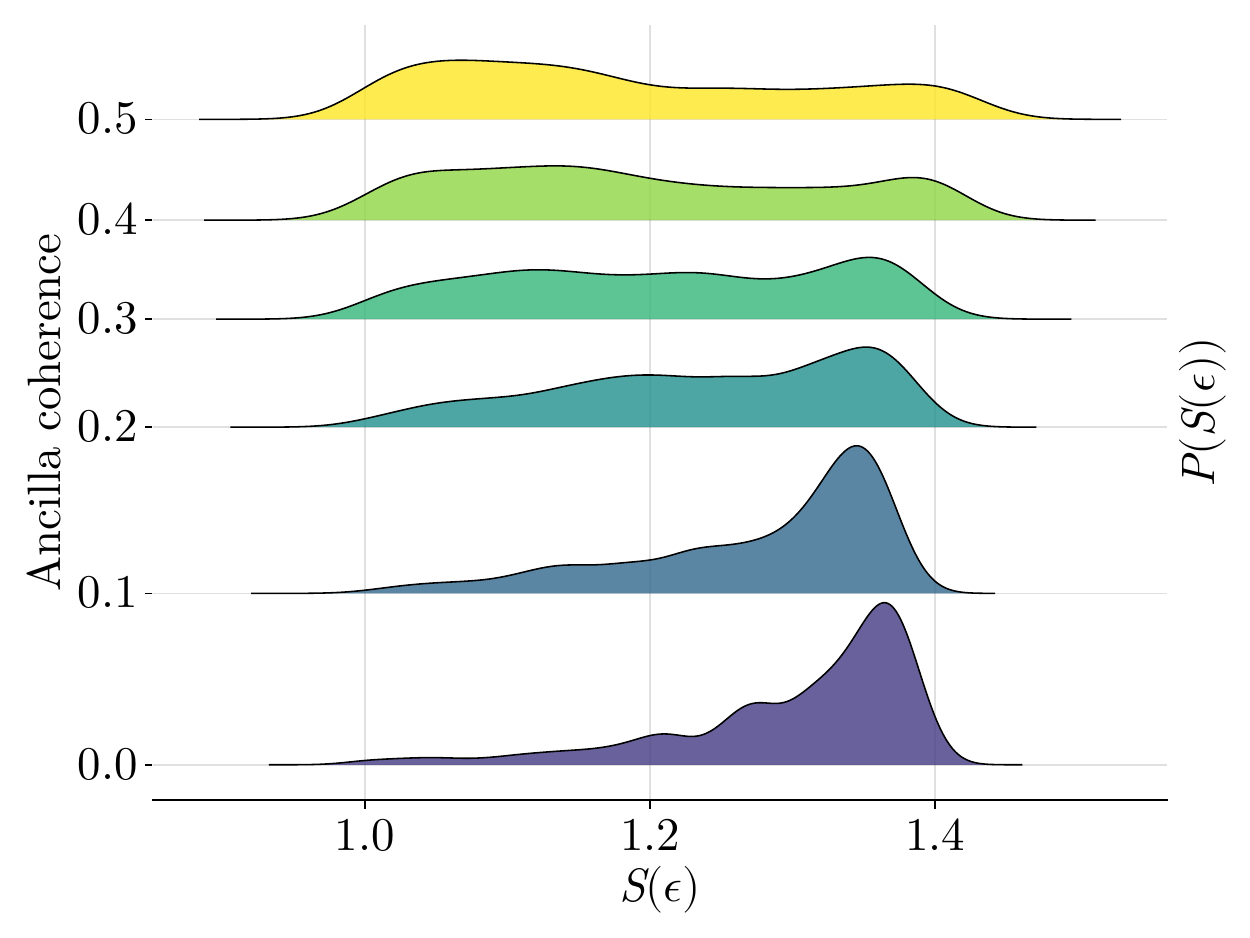}
        \textbf{(b)}
    \end{minipage}
    \caption{Normalized speedup distributions $P(S(\epsilon))$ over $1000$ $q_1$ random Haar states, as a function of (a) the ancilla qubit's excited state population (incoherent states) and (b) the ancilla's coherences, with a ground state population fixed to $0.5$.
    }
    \label{fig:Speedup_distributions}
\end{figure}

To gain further insight into the distribution of speedups, we next vary the ground-state population $p_2^{0}$ of the ancilla qubit from $0$ to $1$ in steps of $0.2$, while keeping the ancilla state diagonal.
The resulting speedup distributions, obtained from $1000$ Haar-random initial states of $q_1$, are shown in panel~(a) of~\cref{fig:Speedup_distributions}.
We observe that the mean of the distribution remains approximately constant across all values of $p_2^{0}$.
In contrast, the width of the distribution increases monotonically as the initial ancilla excited state population approaches zero.
This broadening reflects an increased sensitivity of the speedup to the specific initial state of $q_1$, even though the average performance of the protocol remains unchanged.

Finally, and most importantly, we examine the impact of coherences in the ancilla state.
To this end, we fix the ancilla populations to $p_2^{0} = p_2^{1} = 1/2$ and vary the coherence
$C = |\langle 0 | \hat{\rho}_2 | 1 \rangle|$
from $C = 0$ up to its maximal allowed value $C = 1/2$.
The corresponding speedup distributions are shown in panel~(b) of~\cref{fig:Speedup_distributions}.
As the magnitude of the ancilla coherence increases, the weight of the speedup distribution progressively shifts towards $S = 1$, corresponding to no speedup.
For small to moderate coherences ($C \leq 0.2$), a pronounced peak around $S \approx 1.35$ remains visible, indicating that the protocol still yields a substantial speedup for a significant fraction of initial states.
For larger coherences, however, the distribution becomes increasingly concentrated near unit speedup.
This behavior reflects the fact that residual coherences in the ancilla prevent the C-$R_y(\pi)$ gate from fully suppressing the system qubit’s overlap with the slowest-decaying Liouvillian mode.
As a consequence, the effectiveness of the protocol becomes strongly dependent on the state of $q_1$.

Overall, these results demonstrate that the protocol is robust against moderate imperfections in the ancilla preparation, but that large ancilla coherences progressively reduce both the magnitude and the reliability of the achievable speedup.

\section{Deriving the non-Markovian effective model}
\label{app:non:markovian}
In this appendix, we derive the effective master equation considered in~\cref{sec:non-markovian:case}. It describes the temporal evolution of a single qubit (the system $S$) subject to amplitude damping at a rate $\Gamma_1$ and phase damping at a rate $\Gamma_\phi/2$, coupled with a strength $\nu_{zx}$ to a TLS (the bath $B$), itself subject to amplitude damping at a rate $\kappa$.
Since the two qubits $q_1$ and $q_2$ evolve independently [Eq.~(\ref{eq:non-Markovian:model})], for simplicity here we consider only the dynamics of a single qubit coupled to one TLS
\begin{equation}
    \mathcal{L}\left[\hat\rho\right] = - i[\hat H,\hat\rho] + \Gamma_1 \mathcal{D}_{\hat\sigma^{q}_-}\left[\hat\rho\right]
    + \frac{\Gamma_\phi}{2} \mathcal{D}_{\hat\sigma^{q}_z}\left[\hat\rho\right]
    + \kappa \mathcal{D}_{\hat\sigma^\mathrm{TLS}_-}\left[\hat\rho\right],
\end{equation}
where $\hat{H} = \hat{H}^{q}_0 + \hat{H}^{\mathrm{TLS}}_0 + \hat{H}_I = \omega_q\hat\sigma^{q}_z + \omega_t \hat\sigma^{\mathrm{TLS}}_z + \nu_{zx} (\hat\sigma^{q}_z \otimes \hat\sigma^{\mathrm{TLS}}_x)$.
In the interaction picture with respect to $\hat{H}_I$, the time evolution of the state of the system is described by
\begin{equation}
    \frac{d}{dt}\hat{\rho}_s(t) = -\int_0^t ds \mathrm{Tr}_B[\hat{H}_I(t),[\hat{H}_I(s), \hat{\rho}(s)]].
    \label{eq:VonNeumann_int}
\end{equation}
Here $\hat{H}_I(t) = \nu_{zx}(\hat{\sigma}_z^{q}(t) \otimes \hat{\sigma}^{\mathrm{TLS}}_x(t))$ where $\hat{O}(t) = e^{i \hat H_I t}\hat O e^{-i \hat H_I t}$ denotes an operator $\hat{O}$ in the interaction picture. The time evolution of the operators $\hat{\sigma}_z^{q}$ and $\hat{\sigma}^{\mathrm{TLS}}_x = \hat{\sigma}^{\mathrm{TLS}}_+ + \hat{\sigma}^{\mathrm{TLS}}_-$ can be obtained from the solutions of the following adjoint master equation for the qubit or the TLS
\begin{equation}
    \begin{aligned}&\dot{\hat{\sigma}}^{\mathrm{TLS}}_+ = i[\hat{H}_0^{\mathrm{TLS}}, \hat{\sigma}_+^{\mathrm{TLS}}] + \kappa \mathcal{D}_{\hat{\sigma}_-^{\mathrm{TLS}}}^\dagger[\hat{\sigma}_+^{\mathrm{TLS}}] = -(2i\omega_t + \frac{\kappa}{2})\hat{\sigma}_+^{\mathrm{TLS}},\\
&\dot{\hat{\sigma}}^{\mathrm{TLS}}_-= (2i\omega_t - \frac{\kappa}{2})\hat{\sigma}_-^{\mathrm{TLS}},\\
&\dot{\hat{\sigma}}^{q}_z = i[\hat{H}_0^{q}, \hat{\sigma}_z^{q}] + \frac{\Gamma_\phi}{2} \mathcal{D}_{\hat{\sigma}_z^{q}}^\dagger[\hat{\sigma}_z^{q}] = 0 \,,
    \end{aligned}
\end{equation}
where we neglected the intrinsic damping with a rate $\Gamma_1$ of the qubit system. This is justified if we assume that the intrinsic relaxation dynamics of the qubit is slower than the one provided by its interaction with the TLS. This yields
\begin{equation}
    \begin{aligned}\hat{\sigma}^{\mathrm{TLS}}_+(t) &= e^{-(2i\omega_t + \frac{\kappa}{2})t} \hat{\sigma}^{\mathrm{TLS}}_+(0), \\
        \hat{\sigma}^{\mathrm{TLS}}_-(t) &= e^{(2i\omega_t - \frac{\kappa}{2})t} \hat{\sigma}^{\mathrm{TLS}}_-(0), \\
       \hat{\sigma}^{\mathrm{q}}_z(t) &= \hat{\sigma}^{\mathrm{q}}_z(0) \,. \\
    \end{aligned}
\end{equation}
Expanding~\cref{eq:VonNeumann_int} yields four terms
\begin{equation}
    \begin{aligned}
        \frac{\mathrm{d} \hat{\rho}_s}{\mathrm{d}t} = -\int_0^t ds \mathrm{Tr}_B\left[\hat{H}_I(t)\hat{H}_I(s)\hat{\rho}(s) - \hat{H}_I(t)\hat{\rho}(s)\hat{H}_I(s) - \hat{H}_I(s)\hat{\rho}(s)\hat{H}_I(t) + \hat{\rho}(s)\hat{H}_I(s)\hat{H}_I(t)\right].
    \end{aligned}
    \label{eq:VonNeumann_dev}
\end{equation}
We can then perform the Born approximation $\hat{\rho}(s) \approx \hat{\rho}_s(t) \otimes \ket{0}\!\bra{0}$
which makes it possible to make appear the bath correlation function
\begin{equation}
    \begin{aligned}
        \mathrm{Tr}_B(\hat{\sigma}_-^{\mathrm{TLS}}(t)\hat{\sigma}_+^{\mathrm{TLS}}(s)\ket{0}\!\bra{0}) &= \expval{\hat{\sigma}_-^{\mathrm{TLS}}(t)\hat{\sigma}_+^{\mathrm{TLS}}(s)} = e^{-\frac{\kappa}{2}(t+s)}e^{2i\omega_t(t-s)}.
    \end{aligned}
\end{equation}
Inserting this into Eq.~(\ref{eq:VonNeumann_dev}) yields the Redfield master equation~(\ref{eq:reduced_liouv}) of the main text, i.e.,
\begin{equation}
    \begin{aligned}
        \frac{\mathrm{d} \hat{\rho}_s}{\mathrm{d}t} &= \mathcal{L}_{\mathrm{red}}\left[ \hat{\rho}_s\right] = -i[\omega_q\hat{\sigma}_z, \hat{\rho}_s] \\ &+\left(\frac{\Gamma_\phi}{2} + \frac{\nu^2_{zx}}{\frac{\kappa^2}{4}+4\omega^2_t}\left[ -\kappa e^{-\kappa t} + e^{\frac{-\kappa}{2} t} \left( \kappa \cos(2\omega_t t) + 4\omega_t\sin(2\omega_t t) \right)\right] \right)\mathcal{D}_{\hat{\sigma}_z}[\hat{\rho}_s]  + \Gamma_1\mathcal{D}_{\hat{\sigma}_-}[\hat{\rho}_s]. \\
    \end{aligned}
    \label{eq:reduced_liouv_app}
\end{equation}
Solving this system of coupled differential equations yields
\begin{equation}
    \hat{\rho}_s(t) = 
    \begin{pmatrix}
        \rho_{00}(0) + \rho_{11}(0)(1-e^{-\Gamma_1t}) & \Lambda \rho_{01}(0) \\
        \Lambda^*\rho_{10}(0) & e^{-\Gamma_1t}\rho_{11}(0)
    \end{pmatrix}\,,
\end{equation}
where 
\begin{equation}
    \Lambda(t) = \exp(-\frac{16\left(1+e^{-t \kappa}\right) v^2+t(-4 i \delta +\Gamma_1+2 \Gamma_\phi+4 i \omega_t)\left(K^2+16 \omega_t^2\right)-32 e^{-\frac{t \kappa}{2}} \nu_{zx}^2 \operatorname{cos}(2 t \omega_t)}{2\left(\kappa^2+16 \omega_t^2\right)}),
\end{equation}
and $\delta = |\omega_q-\omega_t|$ is the detuning between the qubit and the TLS.
From this, one can obtain an analytical expression of the trace distance to the steady state
\begin{equation}
D(\hat{\rho}_s(t), \ket{0}\!\bra{0}) = \sqrt{|\Lambda(t)\rho_{01}(0)|^2 + e^{-2\Gamma_1t}\rho_{11}^2(0)}.
\end{equation}

\FloatBarrier
\begin{sloppypar} %TO AVOID LONG URL IN REF 2 BREAKING THE MARGIN
\printbibliography
\end{sloppypar}

\end{document}